\documentclass{aa}
\usepackage[varg]{txfonts}
\usepackage{natbib}
\usepackage{graphicx}
\usepackage{placeins}
\usepackage{stfloats}
\usepackage{float}

\bibpunct{(}{)}{;}{a}{}{,}
\usepackage[]{hyperref} 
\hypersetup{
    colorlinks,
    linkcolor=magenta,
    citecolor=blue,
    urlcolor=red}
\usepackage{color}

\begin{document}

\title{A formation pathway for terrestrial planets with moderate water content involving atmospheric-volatile recycling}

\author{Jonas Müller\inst{1,2,4}
\and Bertram Bitsch\inst{3,4}
\and Aaron David Schneider\inst{5,6}}

\titlerunning{A formation pathway for terrestrial planets with moderate water content involving atmospheric-volatile recycling}
\authorrunning{J. M\"uller et al.}

\institute{Heidelberger Institut für Theoretische Studien, Schloss-Wolfsbrunnenweg 35, 69118 Heidelberg, Germany
\and Zentrum für Astronomie (ZAH/LSW), Heidelberg University, Königstuhl 12, 69117 Heidelberg, Germany
\and Department of Physics, University College Cork, Cork, Ireland
\and Max-Planck-Institut für Astronomie, Königstuhl 17, 69117 Heidelberg, Germany
\and Centre for ExoLife Sciences, Niels Bohr Institute, Øster Voldgade 5, 1350 Copenhagen, Denmark
\and Instituut voor Sterrenkunde, KU Leuven, Celestijnenlaan 200D, 3001 Leuven, Belgium}

\date{Received <date> /Accepted <date>}


\abstract{
Of the many recently discovered terrestrial exoplanets, some are expected to harbor moderate water mass fractions of a few percent. The formation pathways that can produce planets with these water mass fractions are not fully understood.
Here, we use the code \texttt{chemcomp}, which consists of a semi-analytical 1D protoplanetary disk model harboring a migrating and accreting planet, to model the growth and composition of planets with moderate water mass fractions by pebble accretion in a protoplanetary disk around a TRAPPIST-1 analog star. This star is accompanied by seven terrestrial planets, of which the outer four planets likely contain water mass fractions of between 1\% and 10\%.
We adopt a published model that considers the evaporation of pebbles in the planetary envelope, from where recycling flows can transport the volatile vapor back into the disk.
We find that with this model, the planetary water content depends on the influx rate of pebbles onto the planet. A decreasing pebble influx with time reduces the envelope temperature and consequently allows the formation of planets with moderate water mass fractions as inferred for the outer TRAPPIST-1 planets for a number of different simulation configurations. 
This is further evidence that the recycling of vapor is an important component of planet formation needed to explain the vast and diverse population of exoplanets.
}

\maketitle

\section{Introduction}

Over recent years, the number of observed exoplanets has steadily increased. This number recently passed the 5000 mark, and is very likely to become even larger in the future. The population of planets and exoplanets can be roughly divided into two groups: terrestrial planets, whose envelope mass is small compared to the mass of their core, and giant planets, whose envelopes have a similar or greater mass than their core.

The bulk water mass fraction stored in the cores of terrestrial planets cannot be obtained directly from a spectrum, but can be inferred using an interior composition model if the mass and the radius ---and hence the density--- of the planet are known \citep[e.g.][]{zeng+2019}. 
In contrast to the relatively small amounts of water found on the terrestrial planets of the Solar System, many terrestrial exoplanets are expected to harbor higher water mass fractions, which can range from a few percent to as high as 50\%, with these latter exoplanets being referred to as water worlds \citep[e.g.,][]{seager+2007, Damasso+2018, tsiaras+2019, benneke+2019}. 

A planetary system of particular interest is the TRAPPIST-1 system, which consists of an early M dwarf star orbited by seven terrestrial planets \citep{gillon+2016, gillon+2017}, whose densities have been estimated by transit and radial velocity observations \citep[e.g.,][]{agol+2021}.
The masses of the planets range from about the mass of Mars to four-thirds the mass of Earth \citep{agol+2021}. \citet{raymond+2022} investigated the water contents of the TRAPPIST-1 planets using five different interior models and assumed uniform rocky interiors among all seven planets given that the relative refractory abundances of planets within one system are expected to be similar \citep[e.g.,][]{dorn+2015, unterborn+2018}, finding that the estimated water mass fractions of the planets range from negligible amounts in the inner system (i.e., for the planets b, c, and d) to up to ${\sim}$10\% for the outer planets \citep[e, f, g, and h; see][]{raymond+2022}.
These planets are prime targets for the \textit{James Webb} Space Telescope (JWST), which may reveal additional constraints on their atmospheric abundances, helping to further constrain the planet formation pathway.

The formation pathways that can produce planets with moderate water mass fractions of several percent are not fully understood. 
An important factor determining the final water mass fraction of a planet is the water abundance of the solids in its vicinity in the protoplanetary disk. Over time, the dust particles in the disk grow into pebbles \citep{brauer+2008} and then into planetesimals \citep[for a review, see][]{johansen+2014}, leading to the formation of planetary embryos that can grow into fully grown terrestrial planets through the accretion of pebbles and planetesimals. Depending on the distance from the star, the pebbles in the disk either contain water ice in the outer regions of the disk (exterior to the water-evaporation line) or contain no water in the inner regions (interior to the water-evaporation line, where water is available only as vapor). However, the pebbles do not remain at a fixed location in the disk, but interact with the gas disk and drift inward \citep{weidenschilling1977, brauer+2008}. When they cross the water-evaporation line, they lose their volatile water contents by evaporation, resulting in a pile-up of icy pebbles outside the water-evaporation line due to the recondensation of outwardly diffusing vapor \citep[e.g., ][]{ros+johansen2013, schoonenberg+ormel2017, drazkowska+alibert2017, schneider+bitsch2021a}.

If the planet forms exterior to the water-evaporation line, its final water mass fraction is expected to be higher than ${\sim}$10\% \citep[e.g., ][]{izidoro+2017, izidoro+2021, izidoro+2022, bitsch+2019b, liu+2019, venturini+2020, schlecker+2021}. On the other hand, if the planet forms within the water-evaporation line, one expects its water content to be negligible. In order to form planets with moderate water mass fractions, some authors have suggested that the TRAPPIST-1 planets begin their growth at the water-evaporation line and then proceed to migrate into the dry inner disk to their current locations \citep{ormel+2017, schoonenberg+2019}. In this way, the planets are theoretically able to accrete both icy and dry material, resulting in an intermediate planetary water mass fraction. However, these works neglected some important processes that might have an impact on the final water content of the planet, such as the outward migration driven by the heating torque \citep{benitez-llambay+2015, masset2017, guilera+2019, baumann+bitsch2020} and the dynamics occurring in the planetary envelope \citep[e.g., ][]{ormel+2015b, lambrechts+lega2017, cimerman+2017, kurokawa+tanigawa2018}.
In addition, the pile-up of icy pebbles outside the water-evaporation front \citep[e.g., ][]{ros+johansen2013, schoonenberg+ormel2017, drazkowska+alibert2017, schneider+bitsch2021a} could significantly increase the water content of the planets forming there, especially considering that the water-evaporation front might harbor a migration trap, and even a pebble trap \citep{müller+2021}.

Hydrodynamical simulations show that the planetary envelope might be penetrated to a certain depth by so-called recycling flows. These recycling flows are potentially able to transport volatile vapor from the envelope back into the disk \citep{ormel+2015b, lambrechts+lega2017, cimerman+2017, kurokawa+tanigawa2018, wang+2023}.
\citet{johansen+2021} demonstrated that this process can tremendously reduce the amount of water that is accreted onto the planet. In their simulations, large fractions of evaporated water ice are recycled instead of being bound to the planet by gravity. 
Using a pebble-accretion model for terrestrial planets, which also includes contributions from planetesimals, \citet{johansen+2021} were able to reproduce the Earth's water content
in their simulations without making the assumption that Jupiter completely blocks the influx of pebbles through the disk, which is a commonly invoked scenario to explain why the inner planets of the Solar System are water-poor \citep[e.g., ][]{morbidelli+2016, kruijer+2017}. 
However, this raises the question of whether the pebble-accretion model that accounts for the recycling flows is also capable of explaining intermediate planetary water fractions of several percent, such as those inferred for the outer TRAPPIST-1 planets.

In this work, we use the semi-analytical code \texttt{chemcomp} \citep{schneider+bitsch2021a}
to study how different assumptions regarding pebble drift through the protoplanetary disk and the opacity of the planetary envelope affect the final water mass fractions of planets.
The code simulates a 1D model of the protoplanetary disk while considering a simple chemical partitioning model of the gas and the solids and the growth of one planet within that disk. We incorporate the treatment of the recycling flows into our model to investigate how pebble evaporation in the disk and the drift behavior of the pebbles through the disk influence the final planetary water content. In addition, we invoke a state-of-the-art envelope opacity model by \citet{brouwers+2021}, which accounts for the growth, fragmentation, and erosion of pebbles during their sedimentation through the planetary envelope, and compare it to the simpler envelope opacity model used by \citet{johansen+2021}, which scales only with the envelope temperature. 
Overall, we adjust the model of \citet{johansen+2021} by introducing more sophisticated prescriptions for the envelope opacity and the pebble drift in order to find a formation pathway for planets with intermediate water mass fractions.

This work is structured as follows. In Sect. \ref{sect: envelope and opacity model}, we describe the setup of our simulations, and the implementation of the recycling flows and the Brouwers opacity into the code. 
In Sect. \ref{sect: influence on the planetary water content}, we discuss the influence of the choice of the envelope opacity model of the planet and the drift model of the pebbles in the disk on the final planetary water content.
In Sect. \ref{sect: modeling the TRAPPIST-1 planets},
we compare the results obtained with the adjusted model of \citet{johansen+2021} with the water mass fractions estimated for the outer TRAPPIST-1 planets.
Finally, we discuss our findings in Sect. \ref{sect: discussion} and summarize our conclusions in Sect. \ref{sect: summary}.

\section{Simulation setup} \label{sect: envelope and opacity model}

\subsection{Initialization} \label{sect: initialization}

The simulations presented in this work were performed using the recently developed and open-source \texttt{chemcomp} code \citep{schneider+bitsch2021a, schneider+bitsch2024}. It solves the equations for the evolution of the dust \citep[][]{birnstiel+2012} and gas \citep[][]{lynden-bell+pringle1974, pringle1981, armitage2013} in a protoplanetary disk harboring a growing and migrating planet in 1D.

The disk quantities are parameterized on a radial logarithmic grid with $N_{\rm grid} = 500$ grid cells. The location of the outer edge corresponds to an orbital distance of $r_{\rm out} = 1000$ AU where the surface density is negligible, and the inner edge is at $r_{\rm in} = 0.01$ AU. The location of the inner edge is motivated by the magnetospheric cavity \citep{bouvier+2007}, in which the magnetic field of the star pulls ionized material away from the midplane of the disk. The material is then funneled onto the star. This is expected to take place near the corotation radius \citep[e.g., ][]{gunther2013}, at which the magnetic field of the star rotates at the same speed as the gas. We stop the simulation at the end of the disk's lifetime or when the planet reaches the inner edge of the disk.

We set the mass of the central star to $M_* = 0.09\ M_\odot$ and the stellar luminosity to $L_* = 0.01\ L_\odot$. The values are based on observational estimates for the stellar mass of TRAPPIST-1 \citep{mann+2019} and the theoretical evolution history of the luminosity \citep[e.g., ][]{baraffe+1998, baraffe+2015}.
We focus on a TRAPPIST-1 analog to allow comparability with the mass fractions of planetary water estimated from observations \citep[e.g., ][]{raymond+2022}.
The initial mass and radius of the protoplanetary disk scale with the mass of the host star \citep[e.g., ][]{andrews+2010, pascucci+2016, ansdell+2017}. We chose an initial disk radius of 
$R_0 = 25$ AU, $35$ AU, or $45$ AU and initial disk masses that range between 3\% and 10\% of the stellar mass of TRAPPIST-1. 
These values of the protoplanetary disk are motivated by using the scaling relation of  \citet{andrews+2010} for mass and radius ($M_0 \propto R_0^{1.6}$). 
A list of the values of the parameters that we use in our simulations is provided in Table \ref{tab: simulation parameters}. Details on the operating principle of the code can be found in \citet{schneider+bitsch2021a}.

In the following, we describe the additions to our planet formation model \citep{schneider+bitsch2021a}, which include recycling flows (Sect. \ref{sect: recycling flows}), a new planetary envelope model (Sect. \ref{sect: structure equations}), and a new envelope opacity model (Sect. \ref{sect: envelope opacity model}).
Apart from the treatment of the planetary envelope, we make the following adjustments to the \texttt{chemcomp} code as it has been presented in \citet{schneider+bitsch2021a}. Firstly, we change the initial mass of the planetary embryo to a fixed value of $M_{\rm t} = 10^{-3}\ M_\oplus$ in order to enable comparability with \citet{johansen+2021}. Secondly, we follow \citet{liu+2019} and insert a linear scaling term that depends on the stellar mass into the formula of the pebble isolation mass used by the code \citep{bitsch+2018}. Thirdly, we replace the accretion luminosity used by the \texttt{chemcomp} code \citep{chrenko+2017,schneider+bitsch2021a} with Eq. \ref{eq: total accretion luminosity} when computing the thermal torque when utilizing the envelope module. Lastly, we neglect gas accretion, type-II migration, and the gap opening of massive planets, since in this work we focus on terrestrial planets for which these processes are not relevant.

\subsection{Planetary envelope model} \label{sect: planetary envelope model}

\subsubsection{Recycling flows} \label{sect: recycling flows}

Planets gain mass through the accretion of solids during the early phases of their growth. In our model, this is realized by the accretion of inward-drifting pebbles that are caught by the planet's slowly increasing gravitational influence on its vicinity \citep[following the prescriptions of][]{johansen+lambrechts2017}. The accretion of pebbles onto the planet releases large fractions of their potential energy. This is expressed by the so-called accretion luminosity (see Eq. \ref{eq: total accretion luminosity}), which heats the growing planet and its envelope. The accretion luminosity determines the thermal structure of the planetary envelope, such that the local envelope temperature, pressure, and density are higher at smaller distances from the surface of the planet. Therefore, the pebbles experience a change in temperature as they are accreted. At some point, the local envelope temperature and pressure might meet the conditions for evaporation of a certain volatile species contained by the pebbles, causing this species to gas out.
In Table \ref{tab: condensation temperatures}, we list all volatile and refractory species present in our model of the protoplanetary disk and their corresponding evaporation temperatures.

It has been shown by hydrodynamical simulations that flows from the protoplanetary disk penetrate the planetary envelope down to a depth at which the local relative entropy of the envelope (see Eq. \ref{eq: relative entropy}) becomes ${\sim}$ 0.2 \citep[][]{ormel+2015b, lambrechts+lega2017, cimerman+2017, kurokawa+tanigawa2018}\footnote{We note that hydrodynamical simulations with a different setup can come to different conclusions \citep[e.g., ][]{moldenhauer+2021, moldenhauer+2022, bailey+2023}.}. The buoyancy barrier prevents the flows from going deeper. This means that if a volatile evaporates at a depth in the envelope where the relative entropy level is 0.2 or higher, the volatile vapor might be recycled back into the disk instead of remaining bound to the planet. Therefore, the flows from the disk are called "recycling flows". The recycling flows transport the vapor back across its corresponding evaporation front where the vapor nucleates on dust particles, since these have the dominant surface area. Decisive for whether the vapor can be recycled back into the disk is whether the coagulation timescale of the ice-covered dust particles is larger than the recycling timescale.
Evaluated at the water-evaporation front in the planetary envelope,
both timescales are generally of the order of several years and are thus comparable \citep[e.g., ][]{johansen+2021, wang+2023}.
However, the coagulation timescale increases strongly exterior to the water-evaporation front due to the decreasing gas density and temperature. Therefore, the freshly nucleated volatile ice particles cannot coagulate sufficiently and are recycled back into the disk \citep{johansen+2021}.

The recycling timescale depends on the Keplerian orbit frequency, which itself depends on the stellar mass and the orbital distance of the planet. 
While \citet{johansen+2021} and \citet{wang+2023} focused on a disk around a Sun-like star, we study planet formation in a disk around a TRAPPIST-1 analog that has less than 10\% of the Sun's mass. The change in stellar mass is roughly compensated by the generally smaller semi-major axis of planets orbiting low-mass stars. Therefore, the timescale for recycling is also on the order of several years, and recycling can also operate in our simulations. However, this does not necessarily apply to planets that grow in the outer disk (i.e., at larger orbital distances), which is why we only consider planets that start their growth in regions where the local disk temperature is equal to or greater than 80 K.

While the accretion of volatiles can be prevented by the recycling flows, the refractory species are always accreted onto the planet in our model. We assume that they are not recycled back into the disk, because they sublimate at greater depths in the envelope compared to the volatiles, at which coagulation is expected to be much more efficient due to higher local dust temperatures and densities. Furthermore, we assume that the vapor of the refractories is deposited at the surface of the planet, where it accumulates and forms an inner high-Z layer, as described by for example \citet{brouwers+ormel2020}. The high-Z layer surrounds the planetary core and consists of vapor from the refractory compounds of incoming pebbles. However, this is important only for the inner regions of the envelope and only when the envelope is hot enough to evaporate the refractories. Since the volatile evaporation fronts, including the water ice line, posses much smaller evaporation temperatures than the refractories, they are located exterior to the high-Z layer. Thus, the high-Z layer does not affect the accretion rates of the volatiles and can be omitted within the scope of this work.

\begin{figure*}[!ht]
    \sidecaption
    \includegraphics[width=12cm]{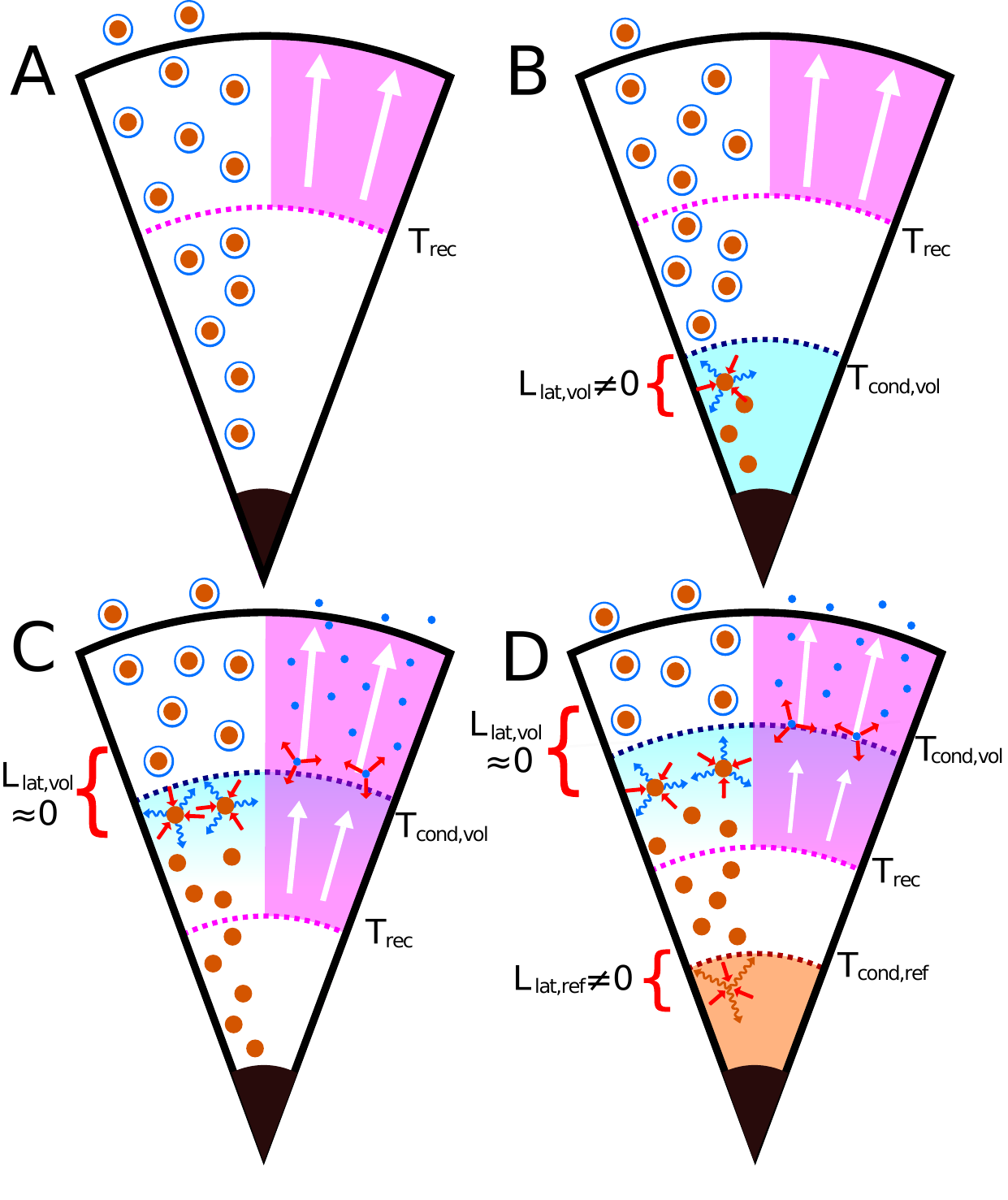}
    \caption{Cartoon of the planetary envelope in different scenarios, each featuring one volatile species (in blue) and one refractory species (in brown). The infalling pebbles are shown on the left side of the envelope. Initially, they consist of a refractory core (brown dot) and an icy volatile layer (blue ring). On the right side, we show the recycling flows (in magenta) and the dust grains (blue dots) that they carry away. The envelope temperature increases overall from A to D. \textit{A:} Cold envelope in which the volatile sublimation temperature is not reached. \textit{B:} The volatile species evaporates in the envelope (indicated by the blue arrows), but the recycling flows do not reach deep enough to prevent the accretion of the volatile. Latent heat is absorbed during the evaporation of the volatile ice (indicated by the red arrows). \textit{C:} The volatile species evaporates in the envelope and the recycling flows transport the vapor back beyond the volatile evaporation front, where the volatile recondenses onto the dust grains. The dust grains covered in ice are then recycled back into the disk. There is no net latent heat absorption due to the recondensation subsequent to the evaporation. \textit{D:} Same as C, but the envelope is hot enough to vaporize the refractory species, causing the absorption of latent heat in the process. The refractory vapor is not recycled.}
    \label{fig: cartoon envelope}
\end{figure*}

In Fig. \ref{fig: cartoon envelope}, we show a cartoon of the envelope of a pebble accreting planet in different scenarios in order to illustrate the recycling of the volatiles.

\subsubsection{Structure equations} \label{sect: structure equations}

We base our model for the planetary envelope and the atmospheric recycling of volatiles on the model of \citet{johansen+2021}. In order to make the model simple and flexible, a number of simplifying assumptions are made. Although we use more sophisticated prescriptions for the pebble drift through the disk \citep{birnstiel+2012} and the envelope opacity profile \citep{brouwers+2021}, many of the simplifications made by \citet{johansen+2021} are also present in our model. The limitations of our model are discussed in Sect. \ref{sect: limitations and caveats}.

Planetary envelopes in the quasi-hydrostatic equilibrium can be characterized by three fundamental structure equations which read \citep[e.g., ][]{kippenhahn+1990}:
\begin{gather}
    \frac{\partial m_{\rm env}}{\partial r_{\rm env}} = 4 \pi r_{\rm env}^2 \rho_{\rm env}
    \label{eq: envelope mass gradient} ,\\
    \frac{\partial P_{\rm env}}{\partial r_{\rm env}} = - \frac{G m_{\rm env} \rho_{\rm env}}{r_{\rm env}^2}
     \label{eq: envelope pressure gradient} ,\\
    \frac{\partial T_{\rm env}}{\partial r_{\rm env}} = \frac{\partial P_{\rm env}}{\partial r_{\rm env}} \frac{T_{\rm env}}{P_{\rm env}} \min ( \nabla_{\rm rad}, \nabla_{\rm ad} ).
    \label{eq: envelope temperature gradient}
\end{gather}
Here, $T_{\rm env}$, $P_{\rm env}$, and $\rho_{\rm env}$ are the gas temperature, the gas pressure, and the gas density of the envelope, $r_{\rm env}$ is the radial distance from the center of the planet, $m_{\rm env}$ is the enclosed mass, and $G$ is the gravitational constant. The temperature gradient in the envelope (Eq. \ref{eq: envelope temperature gradient}) is determined by the Schwarzschild criterion which states that the envelope is convective if the radiative temperature gradient $\nabla_{\rm rad}$ is larger than the adiabatic temperature gradient $\nabla_{\rm ad}$. Convective fluids are well mixed and realize an adiabatic temperature gradient. The radiative and the adiabatic temperature gradients are given by:
\begin{gather}
    \nabla_{\rm rad} = \frac{3 \kappa_{\rm env} L_{\rm tot} P_{\rm env}}{64 \pi \sigma_{\rm SB} G m_{\rm env} T_{\rm env}^4} 
    \label{eq: radiative temperature gradient},\\
    \nabla_{\rm ad} = \frac{\gamma - 1}{\gamma},
    \label{eq: adiabatic temperature gradient}
\end{gather}
with $\kappa_{\rm env}$ being the opacity of the planetary envelope, $\sigma_{\rm SB}$ being the Stefan-Boltzmann constant, $\gamma$ being the adiabatic index, and $L_{\rm tot}$ being the total accretion luminosity of the planet discussed below (Eq. \ref{eq: total accretion luminosity}). 
We use the adiabatic temperature gradient of ideal gas (Eq. \ref{eq: adiabatic temperature gradient}) which does not take phase change processes into account.
In addition to the equations above, the gas temperature, pressure, and density of the envelope can be linked together by assuming an ideal gas:
\begin{gather}
    P_{\rm env}=\frac{\rho_{\rm env} k_B T_{\rm env}}{\mu m_{\rm p}}. 
    \label{eq: ideal gas law}
\end{gather}
Here, $k_B$ is the Boltzmann constant, $\mu$ is the mean molecular weight of the gas, and $m_{\rm p}$ is the proton mass.
As in the model of \citet{johansen+2021}, in which the mean molecular weight is defined as the weighted average of a solar hydrogen-helium mixture and silicate vapor at its saturation vapor pressure, we do not consider changes in $\mu$ caused by the accretion of volatiles for the sake of simplicity. Instead, $\mu$ in our model is always set to the local mean molecular weight of the gas in the disk at the current position of the planet. The mean molecular weight of the gas in the disk dynamically changes with time and orbital distance due to the enrichment of the standard hydrogen-helium mixture in vapor \citep[see][]{schneider+bitsch2021a}. We discuss the limitations arising from this choice in Sect. \ref{sect: limitations and caveats}.

Following \citet{johansen+2021}, the total accretion luminosity can be expressed as
\begin{gather}
    L_{\rm tot} = \frac{G M \Dot{M}_{\rm peb}}{R_{\rm sil}} - Q_{\rm sil} \Dot{M}_{\rm ref} - \sum_{Y_{\rm vol}} (Q_{\rm ice} \Dot{M}_{Y_{\rm vol}}) + L_{26}.
    \label{eq: total accretion luminosity}
\end{gather}
The first term on the right corresponds to the potential energy released by the accretion of the pebbles. The second term represents the latent heat absorbed by the sublimation of the refractories in the envelope. The third term stands for the latent heat absorbed by the sublimation of the volatile ices and the fourth term denotes the luminosity generated by the radioactive decay of $^{26} \text{Al}$. The individual parts and parameters of Eq. \ref{eq: total accretion luminosity} are explained in the following.

The height $R_{\rm sil}$ is either the radius of the planetary core $r_{\rm core}$ (if the refractories do not evaporate in the envelope) or the location of the silicate sublimation front in the envelope (if the refractories do evaporate). The latter is calculated self-consistently by utilizing the expression for the saturated vapor pressure of forsterite (Mg$_2$SiO$_4$) by \citet{lieshout+2014}
\begin{gather}
    P_{\rm vap} = \exp \left( - \frac{65308 \text{ K}}{T_{\rm env}} + 34.1 \right) \text{dyn cm$^{-2}$}.
    \label{eq: silicate sublimation front envelope}
\end{gather}
Following \citet{johansen+2021}, we assume that the evaporation of refractories sets in at a certain pressure.
In our model, the silicates begin to sublimate if the saturated vapor pressure becomes larger than the envelope pressure at a particular depth in the envelope. If the saturated vapor pressure does not reach the envelope pressure throughout the entirety of the envelope, the silicates do not evaporate.
In reality, evaporation does not set in at a certain temperature or pressure, but is a gradual process that depends on the interplay of the local properties of the planetary atmosphere. Overall,
the evaporation of silicates has only a marginal effect on the accretion of volatiles, which evaporate at a much greater distance from the planetary surface due to their lower evaporation temperatures.

We use a saturation equation for the refractories, while we use fixed evaporation temperatures for the volatiles. This choice was made to allow a better comparison with \citet{johansen+2021}, where we follow their evaporation recipe. For the same reason,
we assume that all refractory species considered by the code (evaporation temperature in Table \ref{tab: condensation temperatures} higher than 150 K) behave the same way as the silicates in the envelope. This means that either no refractory species evaporates in the envelope (if $R_{\rm sil} = r_{\rm core}$), or that all refractory species do evaporate (if $R_{\rm sil} > r_{\rm core}$).

When the refractories evaporate in the envelope (i.e., $R_{\rm sil} > r_{\rm core}$), they absorb energy in form of latent heat. Each molecule has a characteristic latent heat value, but for simplicity we assume that all refractory species in our model share the latent heat value which is commonly used for silicates: $Q_{\rm sil} = 7.9 \times 10^{10}$ erg g$^{-1}$ \citep{dangelo+podolak2015}. Due to the absorption of energy, the influence of the latent heat reduces the accretion luminosity and effectively cools the envelope. The refractory pebble accretion rate $\Dot{M}_{\rm ref}$ is the combined mass flux of all refractory species onto the planet.

If the volatiles are recycled back into the disk, their net release of latent heat is expected to be negligible. This is because the sublimation of volatiles is followed by the recondensation of the volatile vapor onto dust particles when they cross their corresponding evaporation front, thereby releasing the latent heat previously absorbed in the sublimation process. However, if a volatile is not recycled and the envelope temperature reaches its evaporation temperature at a certain depth, that volatile vaporizes and latent heat is absorbed in the process. For simplicity, we assume that all volatile species share the latent heat value of water $Q_{\rm ice} =  2.5 \times 10^{10}$ erg g$^{-1}$ \citep{dangelo+podolak2015}, which is multiplied by the individual incoming mass flux $\Dot{M}_{Y_{\rm vol}}$ of each volatile species $Y_{\rm vol}$ in Eq. \ref{eq: silicate sublimation front envelope}.

In reality, the absorption and release of latent heat has a local effect on the structure of the envelope. In our model, the release of latent heat is distributed over the entire envelope and is not localized. This is done for the sake of simplicity and to enable better compatibility with \citet{johansen+2021}, who treat latent heat in a similar (global) manner.

The luminosity generated by the radioactive decay of $^{26} \text{Al}$ is given by \citep{johansen+2021}
\begin{gather}
    L_{26} = r_{26} E_{26} M_{Al} \exp\left( \frac{-t}{\tau_{26}} \right),
\end{gather}
where $r_{26} = \frac{f_{26}}{ m_{26} \tau_{26}}$ is the initial decay rate of $^{26} \text{Al}$ per total aluminum mass, $f_{26} = 1.3 \times 10^{-5}$ is the initial $^{26} \text{Al}$ to $^{27} \text{Al}$ ratio  of the primitive disk material \citep{schiller+2015, larsen+2016}, $m_{26} = 26$ u is the $^{26} \text{Al}$ atom mass, $\tau_{26} = 1.03$ Myr is the decay timescale, $E_{26} = 3.12$ MeV is the energy released per decay \citep{castillo-rogez+2009}, and $M_{Al}$ is the total aluminum mass budget of the planet, which is calculated self-consistently with our chemical partitioning model \citep{schneider+bitsch2021a}. The elemental number ratios in our model are solar \citep{asplund2009}.

The envelope opacity $\kappa_{\rm env}$ is discussed in Sect. \ref{sect: envelope opacity model}.
The implementation of the envelope model and the recycling flows into the \texttt{chemcomp} code is described in Sect. \ref{sect: implementation}. In Sect. \ref{sect: reproduction johansen+2021}, we compare our model to the model of \citet{johansen+2021}.

\subsection{Envelope opacity model} \label{sect: envelope opacity model}

\subsubsection{Simple envelope opacity model}

In their work, \citet{johansen+2021} assumed that the envelope opacity used to calculate the thermal structure of the envelope (see Eq. \ref{eq: radiative temperature gradient}) follows a simple power-law with the temperature:
\begin{gather}
    \kappa_{\rm env} = \kappa_{\rm env,0} \left( \frac{T_{\rm env}}{100 \text{ K}} \right)^{0.5},
    \label{eq: envelope opacity johansen}
\end{gather}
with $\kappa_{\rm env,0} = 0.1$ cm$^2$ g$^{-1}$, $1.0$ cm$^2$ g$^{-1}$, or $10.0$ cm$^2$ g$^{-1}$. As an alternative, we introduce a more sophisticated opacity model in the following section.

\subsubsection{Self-consistent opacity prescription for planetary envelopes} \label{sect: brouwers envelope opacity model}

A recent study by \citet{brouwers+2021} revealed that the envelope opacity of a pebble-accreting planet is not only a function of the temperature, but also crucially depends on the pebble accretion rate and the mass of the planet. Lower planetary masses and high accretion rates lead to higher envelope opacities, causing the envelope to be hotter and predominantly convective. Under these conditions, the size of the pebbles in the envelope is limited by fragmentation and erosion, resulting in a pile-up of pebbles \citep{brouwers+2021}. On the other hand, higher planetary masses and low pebble accretion rates lead to lower envelope opacities, colder temperatures, and a radiatively dominated outer envelope layer. In this case, the pebble sizes are determined by the growth limit \citep{brouwers+2021}.

Both dust and pebbles are present in the envelope of a pebble-accreting planet. If so, both populations must contribute to the total envelope opacity, along with the gas. \citet{brouwers+2021} presented a model describing the sedimentation and production of dust and pebbles in planetary envelopes. It considers the growth (coalescence), fragmentation, and erosion of pebbles.
The total envelope opacity is defined as the sum of the contributions of the pebbles, the dust, and the gas \citep{brouwers+2021}:
\begin{gather}
    \kappa_{\rm env} = \kappa_{\rm peb} + \kappa_{\rm dust} + \kappa_{\rm gas}.
    \label{eq: envelope opacity brouwers}
\end{gather}
The envelope opacity model of \citet{brouwers+2021} is based on two largely unconstrained scaling parameters: the limiting velocity $v_{\rm lim}$ and the fractional dust production efficiency $F$. These parameters and the
calculation of the individual contributions to the total envelope opacity are discussed in Sect. \ref{sect: self-consistent opacity} and Sect. \ref{sect: opacity contribution}. The implementation of the self-consistent opacity model into the \texttt{chemcomp} code is described in Sect. \ref{sect: implementation opacity}.
In the following, we refer to this opacity model as the {\it Brouwers opacity} model.

\subsubsection{Comparison to the simple opacity model} \label{sect: comparison to the simple opacity model}

\begin{figure}[]
    \centering
    \resizebox{\hsize}{!}{\includegraphics{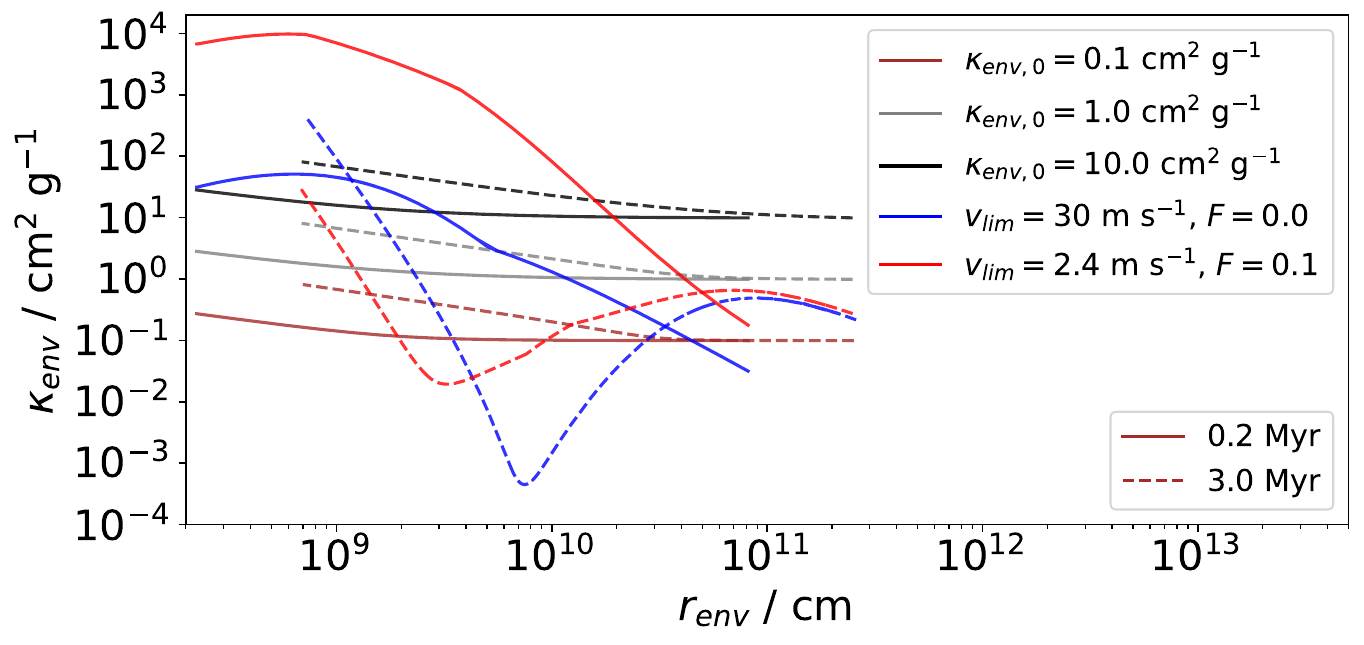}}
    \caption{Brouwers opacity (Eq. \ref{eq: envelope opacity brouwers}) and simple envelope opacity (Eq. \ref{eq: envelope opacity johansen}) as a function of the distance from the center of the planetary core at an early (solid line) and a late stage (dashed line) in the planet's growth. We show the Brouwers opacity that arises when the default parameters of \citet{brouwers+2021} are used ($v_{\rm lim} = 2.4$ m s$^{-1}$, $F = 0.1$), as well as an estimate of the lower limit of the envelope opacity ($v_{\rm lim} = 30.0$ m s$^{-1}$, $F = 0$). The simple opacity is shown for three different opacity scaling factors $\kappa_{\rm env,0}$. The migration of the planet is suppressed, the growth of the planet initiates at $t_0 = 0.1$ Myr, and the disk lifetime is 3 Myr. The incoming pebble flux is fixed to $0.6\ M_\oplus$ Myr$^{-1}$, but can still be lowered by volatile recycling in the planetary envelope. 
    Apart from that, we use our standard parameters (see Table \ref{tab: simulation parameters}). After 0.2 Myr (3.0 Myr), the planets have a mass of about 0.05 $M_\oplus$ (1.5 $M_\oplus$).}
    \label{fig: brouwers opacity comparison}
\end{figure}

In Fig. \ref{fig: brouwers opacity comparison}, we compare the envelope opacities that arise when using the Brouwers opacity model (Eq. \ref{eq: envelope opacity brouwers}) with the opacities that are obtained when adopting the simple opacity model used by \citet{johansen+2021} (Eq. \ref{eq: envelope opacity johansen}). The Brouwers opacities featured in Fig. \ref{fig: brouwers opacity comparison} are the same as the total envelope opacities shown in Fig. \ref{fig: brouwers opacity contributions}.

Compared to the changes in the Brouwers opacity, which can span various orders of magnitude, the simple envelope opacity remains relatively constant in both time and distance from the center of the planet. The simple envelope opacity is a weak function of the envelope temperature only and thus, unlike to the Brouwers opacity, is not directly affected by the increasing mass of the planet. In addition, the simple envelope opacity increases during the growth of the planet, as the planetary envelope is heated by accretion, while the Brouwers opacity decreases with increasing planetary mass \citep{brouwers+2021}. The main reason for this is the inflation of the envelope as the planet grows \citep{brouwers+2021}. We conclude that a simple envelope opacity law that is only a function of the envelope temperature is not able to reproduce key features of the more sophisticated Brouwers opacity model.

\section{Influence on the planetary water content} \label{sect: influence on the planetary water content}

In their simulations, \citet{johansen+2021} found that the planets accrete their water contents early in their growth, until the envelope becomes too hot and water vapor is transported back into the disk by the recycling flows. Here we investigate how the planet's water ice accretion behavior changes as we relax two simplifying assumptions made by \citet{johansen+2021}, namely the simplistic envelope opacity model (Eq. \ref{eq: envelope opacity johansen}) and the assumption that the pebbles in the disk follow the motion of the gas, realizing a pebble surface density that is equal to the gas surface density multiplied by a factor of $\xi_{\rm flux} = 0.0036$. Pebble evaporation and condensation at the evaporation fronts in the disk are neglected in this section, similar to \citet{johansen+2021}.

\subsection{Planetary envelope opacity model} \label{sect: influence of opacity model}

\begin{figure}[]
    \centering
    \resizebox{\hsize}{!}{\includegraphics{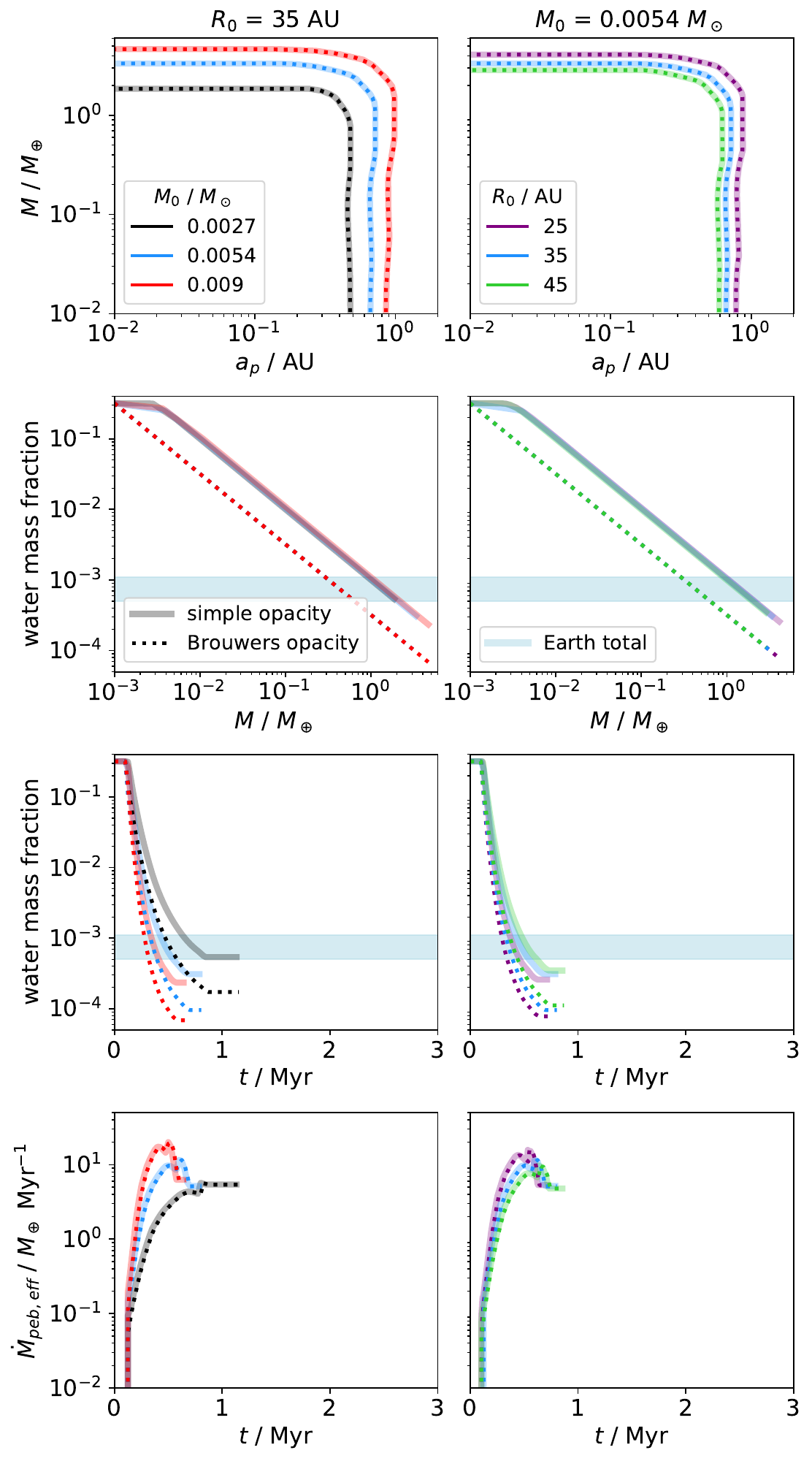}}
    \caption{Comparison of planets using the simple envelope opacity model of \citet{johansen+2021} with $\kappa_{\rm env,0}= 0.1$ cm$^2$ g$^{-1}$ (see Eq. \ref{eq: envelope opacity johansen}; transparent lines) and the Brouwers opacity model by \citet{brouwers+2021} with $v_{\rm lim} = 30.0$ m s$^{-1}$ and $F = 0.0$ (see Eq. \ref{eq: envelope opacity brouwers}; dotted lines) for different initial disk masses {\it (left)} and different initial disk radii {\it (right)}. The pebbles in the disk follow the motion of the gas and behave similarly as in \citet{johansen+2021}. The planets start at a disk temperature of 100 K and migrate toward the inner edge of the disk within 1.5 Myr, causing the simulation to end. The growth of the planet initiates at $t_0 = 0.1$ Myr. The maximum disk lifetime is set to 5 Myr, but for the sake of clarity we only show the first 3 Myr in the bottom two rows.
    The light-blue region corresponds to the estimated total water mass fraction of the Earth, consisting of contributions from the hydrosphere, the exosphere including the crust, and the mantle \citep[e.g., ][]{marty2012, dangelo+2019}. \textit{First row:} Growth tracks. Due to the viscous heating of the protoplanetary disk, the initial position of the planet changes for different initial disk masses and radii, as we want to keep it in the same relative position exterior to the water-evaporation line. \textit{Second row:} Planetary water mass fraction as a function of planetary mass. \textit{Third row:} Planetary water mass fraction as a function of time.  For the sake of clarity, only the first 3 Myr are shown in this and the following row. \textit{Last row:} Effective pebble accretion rates onto the planet as a function of time.}
    \label{fig: johopac}
\end{figure}

To probe the robustness of the water ice accretion behavior found by \citet{johansen+2021} under a change in the envelope opacity model of the planet, we set the pebble surface density equal to the gas surface density multiplied by $\xi_{\rm flux}$ in the simulations presented in this subsection. We still use the chemical compositions calculated by the chemistry model of the \texttt{chemcomp} code (neglecting pebble evaporation and condensation in the disk).
In contrast to the envelope opacity determined with the simple opacity model, the Brouwers opacity tested here tends to decrease with increasing planetary mass. 

In Fig. \ref{fig: johopac}, we compare the water ice accretion behavior of planets using the simple envelope opacity with planets that account for the Brouwers opacity model. 
We recover the early phase of water ice accretion observed by \citet{johansen+2021} when using the simple envelope opacity model (see the second row of Fig. \ref{fig: johopac}).
However, it can be seen that the choice of envelope opacity model has a significant impact on the planetary water content. In particular, the phase of early water ice accretion that is prominent when using the simple envelope opacity model is generally suppressed when the Brouwers opacity is considered. The reason for this is the high initial envelope opacity which enhances the envelope temperature in the early stages of pebble accretion.
This causes the planet to contain significantly less water than when following the simple envelope opacity prescription to the point where it is practically prevented from accreting water at all during pebble accretion.

\subsection{Pebble drift behavior}

\begin{figure}[h!]
    \centering
    \resizebox{\hsize}{!}{\includegraphics{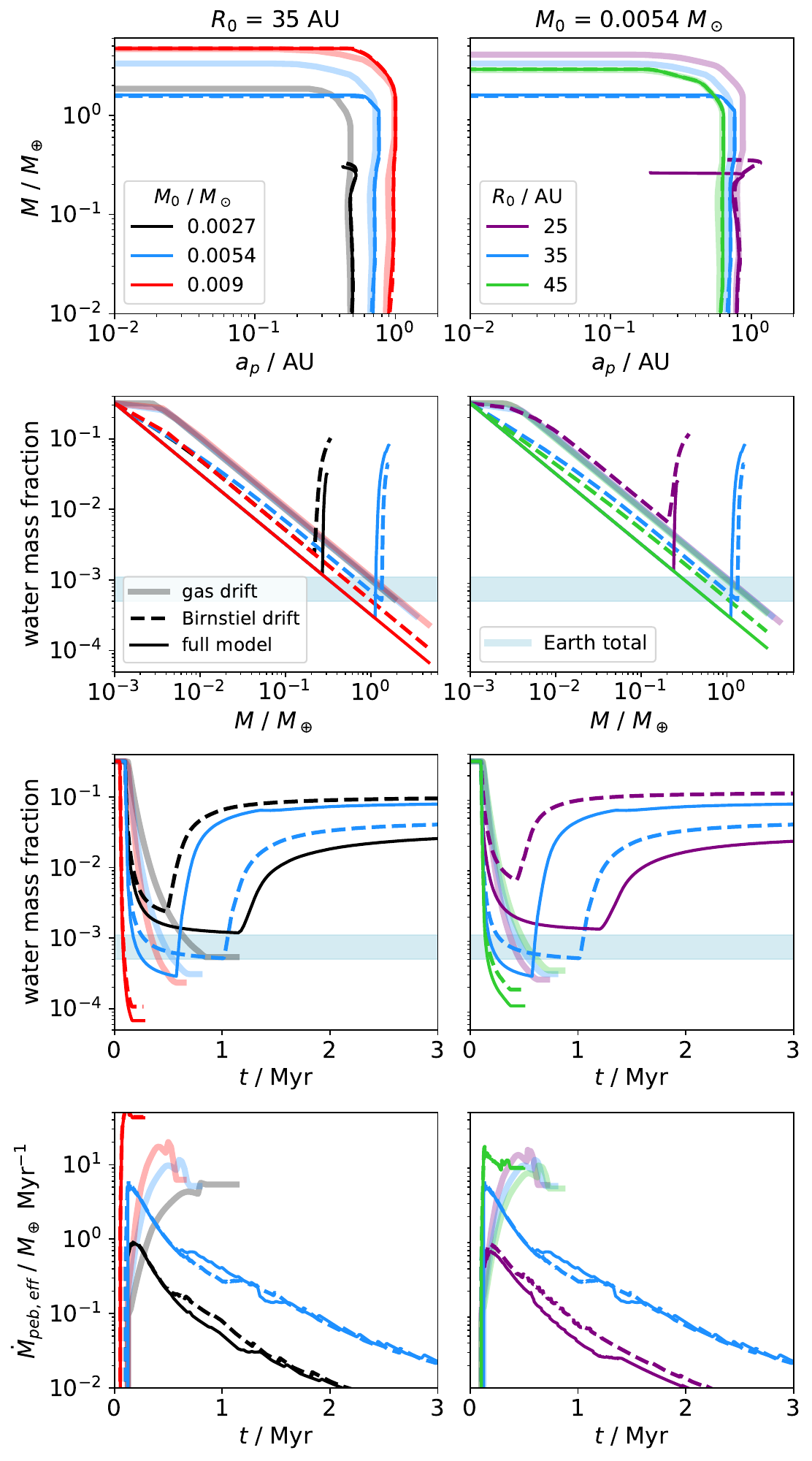}}
    \caption{Comparison of planets growing in a disk with gas-sensitive pebbles (similar to \citet{johansen+2021}; transparent lines) and a disk in which the pebbles follow the Birnstiel pebble drift (dashed lines) for different initial disk masses. The transparent lines are the same as the transparent lines shown in Fig. \ref{fig: johopac}. For these two sets of simulations (transparent and dashed lines), we use the simple envelope opacity model with $\kappa_{\rm env,0}= 0.1$ cm$^2$ g$^{-1}$ (see Eq. \ref{eq: envelope opacity johansen}). 
    In addition, we show simulations using our full model as solid lines. 
    In our full model, the planets grow in a disk with with pebbles drifting according to the Birnstiel pebble drift model while considering the Brouwers opacity model with $v_{\rm lim} = 30.0$ m s$^{-1}$ and $F = 0.0$ (see Eq. \ref{eq: envelope opacity brouwers}). The layout of the plot and the other simulation parameters are similar to Fig. \ref{fig: johopac}.}
    \label{fig: johdrift}
\end{figure}

In Fig. \ref{fig: johdrift} we compare planets that grow in a disk with self-consistently drifting pebbles following \citet{birnstiel+2012}, as implemented in the \texttt{chemcomp} code (see \citet{schneider+bitsch2021a} for details), to planets growing in a disk with gas-sensitive pebbles, as assumed in \citet{johansen+2021} (see Sect. \ref{sect: influence of opacity model}).
In the following, we refer to the pebble drift model as implemented in the \texttt{chemcomp} code as the {\it Birnstiel pebble drift} model.
It can be seen that when the Birnstiel pebble drift is taken into account, the planets in disks with low and medium masses, as well as medium and large initial sizes, end up with considerably higher water mass fractions. This is caused by the facilitated inward drift of pebbles when Birnstiel pebble drift is taken into account, which causes planetary accretion rates to be initially high, but then to steadily decrease due to pebble depletion caused by inward drift \citep[e.g., ][]{birnstiel+2012}. A larger disk mass means that more material is available, which increases the accretion rate on the planet. A smaller disk radius causes the pebbles to grow and drift inwards faster. As a result, the pebble surface density decreases within a few ten thousand years, reducing the accretion rate.

The modified pebble accretion rates result in a change in the water ice accretion behavior of the planets. 
Rather than accreting all their water during the early stages when the envelope is still cold, planets located in disks that are modeled with the Birnstiel pebble drift can accrete water during the later stages, after the influx of pebbles has diminished.
When the incoming pebble flux has decreased sufficiently, the envelope cools, affecting the relative entropy level in the envelope which determines the recycling of volatiles, and the planet is able to accrete water for the remainder of its growth. Since this phase of water ice accretion can last several mega-years, significant amounts of water can be accreted despite the much lower accretion rates of the pebbles in the later phases of the planetary growth. 
The planet growing in the most massive disk migrates very rapidly to the inner edge of the disk, where we stop our simulation. 
Consequently, our simulations do not show the full evolution of the planet. On the other hand, if the planet migrates all the way to the inner disk regions, it will be so hot that it will not accrete water rich material in any case.

When accounting both for the Birnstiel pebble drift and the Brouwers opacity model (solid lines in Fig. \ref{fig: johdrift}), the final planetary water content is higher in some cases, but lower in others than when the simple envelope opacity model is used. Crucial to this is the current mass of the planet as it begins to accrete water ice-rich pebbles. At lower planetary masses, the Brouwers opacity is high, leading to a later onset of water ice accretion than in the case of the simple envelope opacity model. At higher planetary masses, the planet may enter water ice accretion at an earlier time.
This is due to the fact that the Brouwers opacity tends to decrease as the planet becomes more massive, resulting in a higher final planetary water content than when using the simple envelope opacity model.

To sum up, the accretion of icy volatiles is more difficult when the incoming pebble flux is high.
Consequently, a self-consistent incoming pebble flux suppresses water ice accretion at early times but favors it at later times.
In this sense, it has a similar effect on the planet's water ice accretion behavior as the Brouwers opacity model, since the corresponding envelope opacity decreases as the planetary mass increases.

\section{Working example: The TRAPPIST-1 planets} \label{sect: modeling the TRAPPIST-1 planets}

\begin{figure*}[]
    \centering
    \resizebox{\hsize}{!}{\includegraphics{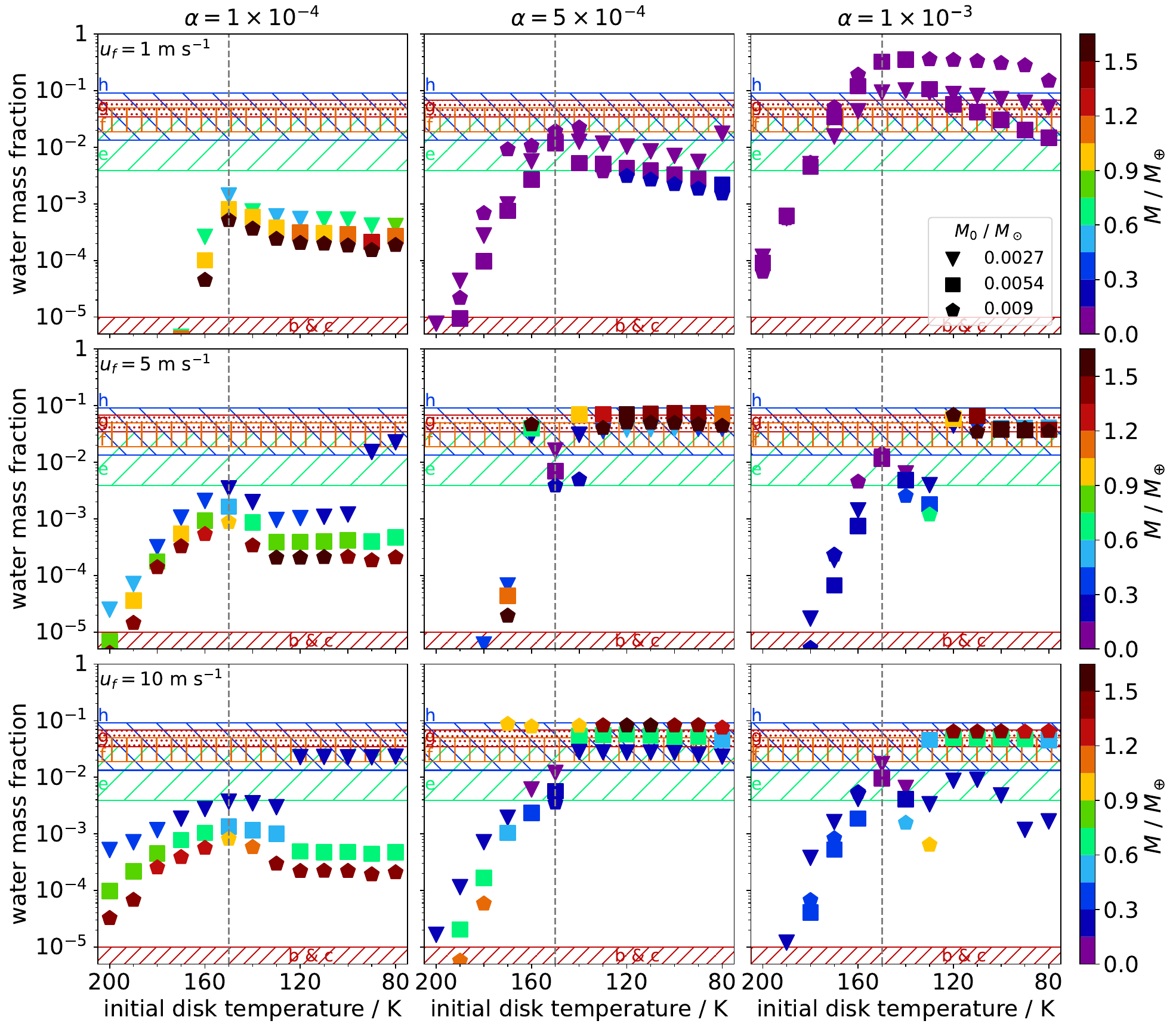}}
    \caption{Final planetary water mass fraction versus initial local disk temperature at the position of the planet as a proxy for the distance from the star.
    Planets that are not shown in the plot have lower water mass fractions than 10$^{-5}$.
    \textit{The columns} correspond to different values of the $\alpha$-viscosity, \textit{the rows} show different solid fragmentation velocity values. Each marker corresponds to one planet starting at a specific disk temperature. The color of the markers indicates the final mass of the planet (see color-bars on the right). The shape of the markers specifies the initial mass of the disk in which the planet grows (see legend).
    The hatched regions in the background correspond to the 1$\sigma$-regions of the water mass fractions of the TRAPPIST-1 planets estimated by \citet{raymond+2022} using their model ii and are color-coded according to the observed masses of the TRAPPIST-1 planets using the same color-bar.
    The vertical lines indicate the location of the water-evaporation front. We use our full model, which consists of both the Brouwers opacity model ($v_{\rm lim} = 30.0$ m s$^{-1}$, $F = 0.0$) and the Birnstiel pebble drift, while applying the planetary envelope model described above. In addition, we account for the planetary migration and the evaporation of pebbles in the disk as they drift inwards.}
    \label{fig: ENVTOT evap}
\end{figure*}

In this section, we compare the planetary water mass fractions obtained in our simulations with the water mass fractions estimated from observations for the TRAPPIST-1 planets \citep[e.g., ][]{raymond+2022}. We use our full model, that is, both the Brouwers opacity model and the Birnstiel pebble drift.

The water mass fraction of the solids in the disk either vanishes interior to the water-evaporation line or lies between ${\sim}$20\% and ${\sim}$30\%\footnote{The water content of the solids exterior to the water-evaporation front in the disk is not constant, as different mixtures of volatile ice are present at different orbital distances (see Table \ref{tab: condensation temperatures}). In addition, it is also influenced by recondensation of volatile vapor onto the solids near the evaporation fronts in the disk \citep[see][]{schneider+bitsch2021a}.} exterior to the water-evaporation line\footnote{The water-evaporation line is located at a disk temperature of 150 K.}. Accounting for the recycling of volatile vapor in the planetary envelope during pebble accretion can significantly reduce the amount of water that is accreted onto the planet \citep{johansen+2021}. Therefore, the final water contents of the planets growing wholly or partially outside the water-evaporation line are expected to be lower when considering volatile recycling.
This means that even a planet accreting all of its mass outside the water-evaporation line could have a water content consistent with the derived water contents of the outer TRAPPIST-1 planets, which have moderate water contents of up to ${\sim}$10\%. This is particularly appealing because in Appendix \ref{sect: no volatile recycling}, we show that these intermediate water contents are most likely not due to the planets having crossed the water-evaporation line during their growth.
Atmospheric-volatile recycling could render the requirement that the planet must accrete both inside and outside the water-evaporation line unnecessary.

For a limited parameter space, we investigate the possibility that the formation of outer TRAPPIST-1 planets could be explained by atmospheric recycling of volatiles that reduces the amount of water accreted on the planet.
The water contents of the inner TRAPPIST-1 planets are consistent with the water content of planets that grow entirely interior to the water-evaporation line, and are therefore not analyzed in detail here.

Here, we also take the evaporation of the pebbles in the disk into account \citep[see][]{schneider+bitsch2021a}. As the pebbles drift inward toward the central star,
the disk's temperature increases. Once they cross the evaporation lines of the molecules
of which they are composed, they release the solid content of that molecule as vapor into the disk.
Due to the outward diffusion and recondensation of the vapor, the pebble surface density and the solid mass fraction of the molecule corresponding to the evaporation line are strongly enhanced near the evaporation line \citep[e.g., ][]{drazkowska+alibert2017, schoonenberg+ormel2017, ros+2019, schneider+bitsch2021a}. The pile-up of water-rich pebbles just outside the water-evaporation line theoretically allows the planets that form there to gain higher water contents than it is possible anywhere in the disks containing nonevaporating pebbles.

In Fig. \ref{fig: ENVTOT evap}, the water mass fraction and the mass of the planet are plotted against its initial position in the disk. Due to the fact that the condensation of water at the water-evaporation line is not instantaneous, low amounts of water ice are present at small distances from the 150 K mark in the disk interior to the water-evaporation line, causing the transition from planets that have a high water content to planets that do not harbor any significant water contents to be smooth.

Overall, the planetary water content tends to decrease with increasing initial disk temperature at the location of the planet within the same disk. This is because planets originating from locations with lower local disk temperatures tend to spend more time exterior to the water-evaporation line due to their larger initial orbital distance, and thus accrete more wet material before they migrate interior to the water-evaporation line \citep[see also][]{bitsch+2019b, schoonenberg+2019}. Even some planets that start their growth interior to the water-evaporation line (i.e., at disk temperature greater than 150 K) are able to accrete significant amounts of water in our simulations. 
This is caused by outward migration via the thermal heating torque which is fueled by pebble accretion and drives the planet to larger orbits in the early phases of its growth, before it migrates back into the inner disk, after it has already accreted most of its final mass. The final masses of the planets depend strongly on the initial mass of the disk, with planets growing in a more massive disk becoming more massive, because more solid material is available.

Fig. \ref{fig: ENVTOT evap} reveals that the recycling of volatiles is indeed efficient in reducing the amount of water ice accreted onto the planet. The final water mass fractions of the planets initiating their growth exterior to the water-evaporation line is relatively robust to changes in the disk mass and starting positions of the planet. The water mass fraction of the solids in the disk exterior to the water-evaporation line
is only reached by the planet in exceptional cases.

The planets residing in a disk with a lower $\alpha$-viscosity generally tend to obtain smaller water mass fractions than the planets growing a disks with a higher $\alpha$-viscosity. This is because the planets in these disks migrate rapidly to the inner edge of the disk, thus preventing them from experiencing a late water ice accretion phase.
In the disks with low $\alpha$, only planets originating from the outer disk are able to accrete a water mass fraction greater than ${\sim} 1\%$, because the local disk temperature serves as a boundary condition for the planetary envelope, with a colder local disk temperature leading to colder envelopes and thus favoring water ice accretion.

For the $\alpha$-viscosity values shown in Fig. \ref{fig: ENVTOT evap}, the planets tend to be the most massive for a solid fragmentation velocity of $u_{\rm f} = 5$ m s$^{-1}$ when compared to simulations with a different value of $u_{\rm f}$ and the same value of $\alpha$. The reason for this is that the pebble size depends sensitively on the fragmentation velocity \citep[][]{birnstiel+2011}. If the fragmentation velocity is 1 m s$^{-1}$, the pebbles are smaller than for $u_{\rm f} = 5$ m s$^{-1}$, resulting in the pebble accretion rate of the planet to be lower. For a fragmentation velocity of 10 m s$^{-1}$, the pebbles are larger, so they drift inwards rapidly and the net pebble accretion rate decreases. Therefore, a fragmentation velocity of 5 m s$^{-1}$ is close to the value at which the highest planetary masses can be reached. The $\alpha$-viscosity also has an effect on the pebble size, but the dependence is less sensitive than for the fragmentation velocity \citep[][]{birnstiel+2011}, causing the observed trend to be present for all values of $\alpha$ tested.

Planets originating from positions corresponding to local disk temperatures near 150 K tend to be less massive than the planets coming from different orbital distances in the same disk. The reason for this is the pile-up of icy pebbles just outside the water-evaporation line, where the local pebble surface density and the water mass fraction of the solids are very high. Due to the high local pebble surface density, the planetary envelope heats up rapidly, causing the accretion of water ice to cease. This greatly lowers the effective pebble accretion rate of the planet, because water is by far the most abundant chemical species at the location of the planet due to the recondensation of water vapor. In addition, the now lower accretion rate triggers the cooling part of the thermal torque, which leads to an early inward migration of the planet.

The pebble-accretion model which accounts for the influence of pebble evaporation in the planetary envelope reliably produces planets with moderate water mass fractions for a wide range of different configurations. 
A planet reaching a water content similar to that anticipated for the outer TRAPPIST-1 planets is the expected result in these disks.
The simulated masses of these planets are also consistent with the mass range of the outer TRAPPIST-1 planets for several configurations, in particular for a viscosity of $\alpha = 5 \times 10^{-4}$.
This is in contrast to the pebble-accretion model neglecting volatile recycling shown in Appendix \ref{sect: no volatile recycling}, where this is only possible in exceptional cases.
This indicates that volatile recycling in the planetary envelope could be a key ingredient to explain the formation of planets with moderate water contents such as in the TRAPPIST-1 system.

\section{Discussion} \label{sect: discussion}

\subsection{Importance of atmospheric-volatile recycling}

Accounting for the recycling of volatiles in the envelope of a pebble-accreting planet has important consequences for the final water content of the planet. While growing in the icy regions of the disk outside the water-evaporation line, the planets reach water mass fractions above 20\% if the effects of the planetary envelope are neglected (see Appendix \ref{sect: no volatile recycling}), but end up with moderate water abundances of several percent when considering volatile recycling. Therefore, the effects of the recycling flows and pebble evaporation in the planetary envelope should not be neglected.

Taking the recycling of the volatiles into account, the final water mass fraction of the planet depends on the envelope opacity and the incoming flux of pebbles, which is determined by the drift behavior of the pebbles in the disk. In general, the self-consistent calculation of both the pebble drift and the envelope opacity leads to the highest final planetary water contents.

Throughout this work, we use relatively low envelope opacities in our simulations to constrain the highest water content a planet can have under certain conditions. However, the actual envelope opacity of a real pebble-accreting planet is not well known. This is reflected by the fact that the scaling parameters of the Brouwers opacity, that is, the limiting velocity $v_{\rm lim}$ and the dust production efficiency $F$, are largely unconstrained \citep{brouwers+2021}. If measurements of these parameters reveal them to be different than the values used here, it is likely that the TRAPPIST-1 planets will end up with smaller water mass fractions than the planets featured in the simulations in Sect. \ref{sect: modeling the TRAPPIST-1 planets}. This could be explored in future simulations.

In this work, we focused on the water content of the planets and have not looked in detail at other volatiles. In particular, we studied the planetary water contents to compare them with the water mass fraction of the TRAPPIST-1 planets estimated in other works \citep{agol+2021, raymond+2022}. However, the other volatiles are also affected by the recycling flows, and the pebble accretion of these volatiles can be prevented by the recycling flows, similar to water, influencing the chemical composition of the planet. As described in Sect. \ref{sect: planetary envelope model}, this is already taken into account by our model and could be investigated in future work. The consideration of volatiles other than water is important especially for planets that form farther away from their host star, at local disk temperatures where most of the volatiles are present in solid form. Therefore, it could have significant consequences, for example, for the composition of the cores of the ice giants in our Solar System.

\subsection{Comparison to the TRAPPIST-1 system}

Using our full model, which includes the recycling of volatiles in the envelope, the Brouwers opacity, the Birnstiel pebble drift, and the evaporation of the pebbles in the disk, the formation of planets with masses and water contents similar to those of the planets f and g is likely for a number of different configurations. Identifying these specific configurations puts constraints on the disk which evolved into the TRAPPIST-1 system as we know it today. According to our simulations, the conditions for the formation of the planets f and g to be possible are a relatively massive disk (about 6\% to 10\% of the stellar mass), a solid fragmentation threshold of 5 m s$^{-1}$ or higher, and a $\alpha$-viscosity of $5 \times 10^{-4}$ or higher. The planetary embryos initiate their growth exterior to the water-evaporation line and migrate inward after accreting most of their final masses.

In our simulations, there is no disk in which planets form that match exactly all four outer TRAPPIST-1 planets.  
This is to be expected due to the simple nature of our model (see also Sect. \ref{sect: limitations and caveats}). Modeling the detailed N-body interactions of the multi-planet TRAPPIST-1 system \citep[see e.g., ][]{papaloizou+2018, huang+ormel2022, teyssandier+2022} is beyond the scope of our work, as well as a more complete treatment of the planetary atmosphere and the recycling flows \citep[see e.g., ][]{ormel+2015b, lambrechts+lega2017, cimerman+2017, kurokawa+tanigawa2018}. Nevertheless, our work indicates that the recycling of volatiles might play an important role in the formation of planets with moderate water contents of a few percent, and in particular in the formation of the outer TRAPPIST-1 planets. 
This opens up new formation pathways that can be further explored in future simulations that address both questions of planetary water content and the architecture of the TRAPPIST-1 system simultaneously.

\subsection{Implications on the formation of the Earth}

In this work, we only attempted to simulate the formation of the TRAPPIST-1 planets. However, by extrapolating our main results to other planets, we can make predictions about the chemical composition and formation history of these planets. It should be noted that we only applied our full model featured in Sect. \ref{sect: modeling the TRAPPIST-1 planets} to disks around a TRAPPIST-1 analog M dwarf star. Disks around more massive star, such as solar-mass stars, are expected to be larger and heavier, shifting the evaporation lines in the disk to greater orbital distances. To accurately model the water content of the Earth, simulations of larger disks are needed that take into account our full model, which is beyond the scope of this work. Nevertheless, since the general thermodynamic structure of the disk is broadly similar, it is justified to make soft predictions for planet formation around other stars based on our results.

Using our full model, the formation of a planet with an Earth-like mass and water mass fraction (i.e., ${\sim} 10^{-3}$) is almost exclusive to the region near the water-evaporation line in a disk with a low $\alpha$-viscosity of $10^{-4}$ (see left column in Fig. \ref{fig: ENVTOT evap})\footnote{Almost exclusive because there is an outlier at 130 K in the disk with $u_{\rm f} = 10$ m s$^{-1}$ and $\alpha = 10^{-3}$.}.
This differs from the model of \citet{johansen+2021}, which assumes that the pebbles in the disk follow the motion of the gas and efficiently produces planets with Earth-like water contents. Accounting for the Birnstiel pebble drift seems to make the formation of these planets more difficult, but not impossible. On the other hand, if a uniform incoming pebble flux were to occur at the position of the growing Earth, perhaps due to the influence of Jupiter, the conditions would be more similar to the model of \citet{johansen+2021} and the formation of an Earth-like planet would be more likely, also for higher $\alpha$-viscosities.

\subsection{Limitations} \label{sect: limitations and caveats}

While the simplicity of our model gives us the advantage of computational efficiency and flexibility over more sophisticated N-body or hydrodynamical simulations, it cannot capture all the physical processes that might play a role in reality. First of all, both our protoplanetary disk model and our planetary envelope model are one-dimensional and cannot account for the 2D and 3D dynamics of the gas, dust and pebbles. Furthermore, our envelope model is a modified version of the envelope model used by \citet{johansen+2021} and inherits a number of limitations from that model. The most important simplifying assumptions and their consequences are discussed below.

As stated in Sect. \ref{sect: structure equations}, the mean molecular weight $\mu$ of the gas in the planetary envelope is set to the local mean molecular weight of the gas in the disk at the current position of the planet. This means that our model does not track the vapor in the envelope and does not account for changes in $\mu$ caused by accretion. This caveat also applies to the model of \citet{johansen+2021}\footnote{Due to its low initial opacity, the planetary envelope is cold enough that the icy pebbles survive the passage to the planetary surface without evaporating in the simulations of \citet{johansen+2021}. In this case, the mean molecular weight of the gas in the envelope should not be significantly affected by accretion. However, this only applies to the early phases of pebble accretion and is no longer the case when volatile recycling begins. The same applies to our simulations with similar assumptions as \citet{johansen+2021}, which are shown in Fig. \ref{fig: johopac} {\it("simple opacity")} and Fig. \ref{fig: johdrift} {\it("gas drift")}.}.
Nevertheless, our model takes into account the change in the chemical composition of the local background gas during the migration of the planet. Accordingly, the mean molecular weight of the gas in the disk considers the pebble drift and the resulting evaporation and recondensation at the evaporation lines in the disk \citep[][]{schneider+bitsch2021a}, which also affects the mean molecular weight in the envelope. The assumption that the mean molecular weight of the envelope is set to the local mean molecular weight in the disk is justified under two conditions:
\begin{enumerate}
    \item Water vapor must be recycled back into the disk. Since water has the highest evaporation temperature of the volatiles in our model (together with H$_2$S, see Table \ref{tab: condensation temperatures}), all volatiles are recycled if water is recycled\footnote{Since our planets begin their evolution at a local disk temperature that is $\geq 80$ K, this is only relevant for H$_2$S and NH$_3$ (which has an evaporation temperature of 90 K).}.
    \item The timescale for filling the recycling layer with volatile vapor must be longer than the corresponding recycling timescale.
\end{enumerate}
If both conditions are met, the volatiles are efficiently recycled and should not significantly affect the mean molecular weight in the envelope. If one or both conditions are not met, volatiles can accumulate in the envelope until they could dominate the mass budget and affect the mean molecular weight. This would lead to a gradient of mean molecular weight in the envelope where $\mu$ increases with depth. In this case, the volatile vapor could form a buoyancy barrier, lowering the local entropy in the envelope and thus preventing the recycling of volatiles altogether. 

Comparing the first condition with our simulations using our full model (i.e., considering the Brouwers opacity and the Birnstiel pebble drift, see Fig. \ref{fig: johdrift}), we find that water is only accreted during the later phases of the planet's growth. This could lead to the formation of an buoyancy barrier in the envelope.
However, since water is accreted until the end of the planet's growth anyway once it has started (which is only possible towards the end of the plant's growth when using our full model), this should not significantly affect the final water mass fraction of the planet.
The same applies to the simulations shown in Fig. \ref{fig: ENVTOT evap}. For other simulations in Figs. \ref{fig: johopac} and \ref{fig: johdrift}, early water accretion occurs,
but as this is not the case for our full model, it is not investigated further of this work.

Since we do not track the volatile vapor in our envelope model, we cannot quantitatively verify the second condition. 
Instead, we perform order-of-magnitude estimates by comparing the recycling timescale of water vapor with the timescale for filling the recycling layer with water vapor. If the latter is larger than the former, our assumption regarding the mean molecular mass is justified. We estimate the timescale for filling the recycling layer with water vapor as the ratio of the current mass of the gas contained in the recycling layer to the water accretion rate (i.e., $\tau_{\rm layer} = M_{\rm layer} / \Dot{M}_{\text{H}_2\text{0}}$). We define the recycling layer of water vapor as the region of the planetary envelope that lies between the water-evaporation line at $T_{\rm env} = 150$ K and the Fe$_3$O$_4$ evaporation line at $T_{\rm env} = 371$ K (or the surface of the protoplanet if this temperature is not reached in the envelope).
These estimates suggest that the timescale for filling the recycling layer with volatile vapor is shorter than the timescale for recycling at very low planetary masses, but becomes longer as the planet grows. This is due to the inflation of the envelope with increasing mass of the planet and the reduction of the influx of pebbles when using the Birnstiel pebble drift model. Therefore, an early phase of water accretion might be possible even when making assumptions similar to our full model. More sophisticated simulations of the planet's atmosphere are needed to assess whether this phase exists and how long it would last. We leave the investigation of this aspect to future work.

Another simplification is that we assume that the recycling flows penetrate the planetary envelope to a depth where the relative entropy reaches 0.2 \citep[similar to][]{johansen+2021}, as found in hyrodynamical simulations of planetary atmospheres \citep{ormel+2015b, lambrechts+lega2017, cimerman+2017, kurokawa+tanigawa2018}. These results were found for a fixed chemical composition of the gas and depend on the setup of the simulations. Hydrodynamical simulations with a different setup come to different conclusions \citep[e.g., ][]{moldenhauer+2021, moldenhauer+2022}.
Furthermore, PdV heating  \citep[i.e., heating by contraction of the accreted gas, see][]{piso+youdin2014} is not being accounted for. It is not expected to be significant because we investigate terrestrial planets undergoing pebble accretion, which prevents substantial gas accretion. Only when pebble accretion seizes can the planet cool and PdV heating can play an important role \citep[for 3D simulations, see][]{schulik+2019}.

Our model shares even more simplifying assumptions with \citet{johansen+2021}, such as the fixed evaporation temperatures of the volatiles, the simple expression for determining the location of the evaporation front of the silicates (Eq. \ref{eq: silicate sublimation front envelope}), the global treatment of latent heat described in Sect. \ref{sect: structure equations}, and the fact that we use the adiabatic temperature gradient of ideal gas which does not take phase change processes into account (Eq. \ref{eq: adiabatic temperature gradient}).
However, we do not expect these simplifications to change our main finding that a late phase of water accretion can occur when the pebble accretion rate of the planet decreases with time.

In Figs. \ref{fig: johopac} and \ref{fig: johdrift}, the initial position of the planet is set to the orbital distance at which the local disk temperature is 100 K. In Fig. \ref{fig: ENVTOT evap} we show the final water content of planets originating from a range of different initial disk temperatures. In this context, it should be noted that the water-evaporation front (which is located at 150 K in our disk model) has been proposed as the first site of planet formation \citep[e.g., ][]{weidenschilling1977, ros+johansen2013, ros+2019, müller+2021}.
Assuming that planets form only at an initial local disk temperature of 150 K, both our full model (see Fig. \ref{fig: ENVTOT evap}) and the model that neglects the atmospheric sublimation of pebbles and volatile recycling (see Fig. \ref{fig: NOENV evap}) are unable to produce planets with masses and water contents in agreement to those of the outer TRAPPIST-1 planets. However, if a slightly wider range of local initial disk temperatures is considered, our model is more likely to produce such a planet, showing that our model can be applicable to more general initial conditions.

\section{Summary} \label{sect: summary}

In this work, we performed semi-analytical 1D simulations of the protoplanetary disk harboring a migrating and pebble-accreting planet using the code \texttt{chemcomp} developed by \citet{schneider+bitsch2021a}. Following \citet{johansen+2021}, we introduced the simplistic treatment of evaporating pebbles within the planetary envelope, as well as the so-called recycling flows found in hydrodynamical simulations \citep[e.g., ][]{ormel+2015b, lambrechts+lega2017, cimerman+2017, kurokawa+tanigawa2018}. These flows are able to transport volatile vapor from the planetary envelope back into the disk, reducing the effective volatile accretion rates of the planet. 

When not accounting for the recycling flows, planets growing exterior to the water-evaporation front typically end up with high water mass fractions of greater than 20\% (see Appendix \ref{sect: no volatile recycling}). However, the water mass fractions of these planets can be drastically reduced if the recycling of the volatiles in the planetary envelope is considered. Making assumptions similar to \citet{johansen+2021}, namely a simple envelope opacity law that scales with the envelope temperature, and that the pebble surface density of the disk is a fraction of the gas surface density, the planets only accrete small amounts of water (less than 1\%). This raises the question how planets with an intermediate water mass fraction of several percent, as inferred for the outer planets of the TRAPPIST-1 system, could have formed.

In our simulations, the key to forming planets with intermediate water mass fractions is taking into account the Birnstiel pebble drift. 
In this way, the planetary accretion rate steadily decreases with increasing planetary mass as the reservoir of pebbles remaining in the disk is depleted by the inward drift of the pebbles. This allows the envelope to cool as the planet grows, and its relative entropy profile, which governs the recycling flows, changes until the accretion of water becomes possible. The water mass fraction of the planet can increase up to ${\sim}$10\% during this late water ice accretion phase, although the exact number depends on the input parameters of the planet and the disk.

Using the Brouwers opacity model, water ice accretion is generally suppressed in the early phases of the planet's growth, but is favored in the late phases. As a result, the early phase of water accretion is completely suppressed, while the occurrence of the late phase of water accretion does not appear to be affected (see Figs. \ref{fig: johopac} and \ref{fig: johdrift}). This indicates that the exact pebble flux is more important than the choice of the envelope opacity model for the late phase of water accretion. However, the onset of the late phase of water accretion can be shifted to either an earlier or later time by using the Brouwers opacity model, depending on the current mass of the planet. This can increase or decrease the final water content of the planet.
Adopting the full model, which also includes pebble evaporation and the recondensation of volatile vapor in the disk, our simulations are able to give birth to planets with masses and water content similar to those of the outer TRAPPIST-1 planets.
More sophisticated simulations are required to check whether this statement is still true if our simplifying assumptions are relaxed (see Sect. \ref{sect: limitations and caveats}).
Overall, our simulations show that accounting for the recycling of volatiles in the planetary envelope while considering a pebble drift that reduces in time could provide a formation pathway for terrestrial planets with moderate water mass fractions via a late phase of water accretion during the disk's lifetime.

\begin{acknowledgements}
J.M. acknowledges funding from the ERC Consolidator Grant Dipolar-Sound (grant agreement \# 101000296). B.B. thanks the European Research Council (ERC Starting Grant 757448-PAMDORA) for their financial support and acknowledges the support of the DFG priority program SPP 1992 ``Exploring the Diversity of Extrasolar Planets'' (BI 1880/3-1). A.D.S. acknowledges funding from the European Union H2020-MSCA-ITN-2019 under Grant no. 860470(CHAMELEON) and from the Novo Nordisk Foundation Interdisciplinary Synergy Program grant no. NNF19OC0057374. We thank Saskia Hekker for her comments that helped improve the manuscript. We also thank the anonymous referee for their report which helped to improve the quality of the manuscript.
\end{acknowledgements}

\bibliographystyle{aa}
\bibliography{ref}

\begin{appendix}

\section{Tables}

Here we provide a table showing the molecules considered in our model and their corresponding condensation temperatures (Table \ref{tab: condensation temperatures}), as well as a table showing the important parameters used throughout this work (Table \ref{tab: simulation parameters}).

\begin{table}[h!]
    \caption{Condensation temperatures of the molecules.}
    \centering
    \begin{tabular}{c|c}
    \hline\hline
    Species (Y) & $T_{\rm cond}$ / K \\
    \hline
    CO & 20 \\
    N$_2$ & 20 \\
    CH$_4$ & 30 \\
    CO$_2$ & 70 \\
    NH$_3$ & 90 \\
    H$_2$S & 150 \\
    H$_2$O & 150 \\
    \hline
    Fe$_3$O$_4$ & 371 \\
    FeS & 704 \\
    NaAlSi$_3$O$_8$ & 958 \\
    KAlSi$_3$O$_8$ & 1006 \\
    Mg$_2$SiO$_4$ & 1354 \\
    Fe$_2$O$_3$ & 1357 \\
    VO & 1423 \\
    MgSiO$_3$ & 1500 \\
    Al$_2$O$_3$ & 1653 \\
    TiO & 2000 \\
    \hline
    \end{tabular}
    \tablefoot{The condensation temperatures of the molecules considered in the simulations are taken from \citet{lodders2003}. The condensation temperature of Fe$_2$O$_3$ is assumed to be equal to the condensation temperature of pure iron. We refer to the molecules in the upper section as volatiles, while those in the lower section are called refractories.}
    \label{tab: condensation temperatures}
\end{table}

\begin{table*}[b]
    \caption{Simulation parameters.}
    \centering
    \begin{tabular}{c|c|c}
    \hline\hline
    Parameter & Value(s) & Meaning  \\
    \hline
    \multicolumn{3}{c}{Planet} \\
    \hline
    $T_{a_{p,0}}$ & 80 K - 200 K, [100 K] & disk temperature at the initial position \\
    $t_0$ & 0.05 Myr & starting time \\
    $\rho_{\rm core}$ & 5.5 g\ cm$^{-3}$ & density of the planetary core \\
    \hline
    \multicolumn{3}{c}{Grid} \\
    \hline
    $r_{\rm in}$ & 0.01 AU & inner edge \\
    $r_{\rm out}$ & 1000 AU & outer edge \\
    $N_{\rm Grid}$ & 500 & number of grid-cells \\
    \hline
    \multicolumn{3}{c}{Disk} \\
    \hline
    $\alpha$ & $1\times10^{-4}$, [$5\times10^{-4}$], $1\times10^{-3}$ & viscous turbulence parameter \\
    $\alpha_{\rm z}$ & $1 \times 10^{-4}$ & vertical mixing parameter \\
    $M_*$ & $0.09\ M_\odot$ & host star mass \\
    $L_*$ & $0.01\ L_\odot$ & host star luminosity \\
    \text{[Fe/H]} & 0.04 & host star metallicity \\
    $M_0$ & 0.0027 $M_\odot$, [0.0054 $M_\odot$], 0.009 $M_\odot$ & initial disk mass \\
    $R_0$ & 35 AU & initial disk radius \\
    $t_{\rm evap}$ & 7 Myr & disk lifetime \\
    $\epsilon_0$ & 2\% & initial solid to gas ratio \\
    $u_{\rm f}$ & 1 m s$^{-1}$, [5 m s$^{-1}$], 10 m s$^{-1}$ & solid fragmentation velocity \\
    \hline
    \multicolumn{3}{c}{Integration} \\
    \hline
    $\Delta t$ & 10 yr & integration time step \\
    \hline
    \end{tabular}
    \tablefoot{The parameters are used for the simulations in this work unless specified otherwise. When multiple values are given for a single parameter, our standard value is bracketed.}
    \label{tab: simulation parameters}
\end{table*}

\FloatBarrier

\section{Volatile recycling in the planetary envelope}

\subsection{Implementation} \label{sect: implementation}

The properties of the planetary envelope are defined on a 1D logarithmic grid with $N_{\rm Grid,env} = 10000$ grid cells. We assume that the envelope is spherically symmetrical. The inner boundary of the grid is equal to the momentary radius of the planetary core $r_{\rm in,env} = r_{\rm core}$ and the outer boundary is set to the current Hill radius of the planet $r_{\rm out,env} = R_H$. 
Therefore, both boundaries are dynamic and change in time while the planet growths. 
In each time step of the simulation, we use the total planetary mass, the disk pressure, and the disk temperature at the location of the planet as boundary conditions ($M$, $P_{\rm disk,env}$, and $T_{\rm disk,env}$) and integrate the Eqs. \ref{eq: envelope mass gradient}, \ref{eq: envelope pressure gradient}, and \ref{eq: envelope temperature gradient} outward in utilizing the solve\_ivp routine (method: ‘RK45’) of scipy package of python \citep{scipy}. As a result, we obtain the profiles of the enclosed mass, the gas pressure, and the gas temperature of the envelope. The envelope density profile is then calculated using Eq. \ref{eq: ideal gas law}.

Similar to the evaporation of the volatiles in the disk, the volatile evaporation in the envelope is governed by fixed evaporation temperatures for simplicity. In reality, the evaporation of a chemical species is not only a function of the temperature, but also depends on other thermodynamic properties such as the pressure. For consistency, we use the same volatile evaporation temperatures in the envelope and in the disk (see Table \ref{tab: condensation temperatures}). This approach causes the accretion of a certain volatile species to be unlikely in the vicinity of its corresponding evaporation front in the disk, because the boundary condition of the envelope temperature is already close to the evaporation temperature of the volatile.

The profile of the relative entropy (i.e., the entropy of the envelope $s_{\rm env}$ divided by the local entropy of the disk $s_{\rm disk}$) is given by
\begin{gather}
    \frac{s_{\rm env}}{s_{\rm disk}} = \frac{P_{\rm env}}{P_{\rm disk,env}} \left( \frac{\rho_{\rm disk,env}}{\rho_{\rm env}} \right)^{\gamma},
    \label{eq: relative entropy}
\end{gather}
where $\rho_{\rm disk,env}$ is the gas density of the disk at the position of the planet. Convective fluids are isentropic. Thus, the relative entropy does only change in the radiatively dominated regions of the envelope. With this, we can introduce the quantity that we call the "recycling temperature $T_{\rm rec}$". It supervises the recycling process and is defined as
\begin{gather}
    T_{\rm rec} = \left\{\begin{array}{ll} \max( T_{\rm env} ), & \min \left( \frac{s_{\rm env}}{s_{\rm disk}} \right) > 0.2\\
    T_{\rm env}\left( \frac{s_{\rm env}}{s_{\rm disk}} \approx 0.2 \right), & \min \left( \frac{s_{\rm env}}{s_{\rm disk}} \right) \leq 0.2\end{array}\right. .
\end{gather}
In the case of $\min \left( \frac{s_{\rm env}}{s_{\rm disk}} \right) \leq 0.2$, $T_{\rm rec}$ is equal to the envelope temperature of the grid cell in which the relative entropy is closest to 0.2.

Since the recycling flows penetrate the envelope to depths where the relative entropy is 0.2, vapor from any depth can be recycled in an envelope where the relative entropy is greater than 0.2 throughout the entire envelope (i.e., $\min \left( \frac{s_{\rm env}}{s_{\rm disk}} \right) > 0.2$). This means that if the envelope temperature reaches the evaporation temperature of a volatile species at any depth, the volatile vapor will be recycled back into the disk instead of being accreted. On the other hand, if the relative entropy becomes lower than 0.2 at a certain depth in the envelope, the volatile vapor is only recycled when its evaporation temperature is reached at a depth where the local relative entropy is still larger than 0.2. Otherwise, the volatile vapor is bound to the planet. For this reason, and because the envelope temperature is a monotonically increasing function with increasing depth in our model, vapor of any volatile is accreted onto the planet only when the recycling temperature $T_{\rm rec}$ is lower than the corresponding evaporation temperature of that volatile.

By comparing $T_{\rm rec}$ with the evaporation temperatures of the volatiles, we can determine whether a volatile is accreted or not.
The potential recycling of volatiles is realized in the code by assigning a factor of either $f_{\rm acc,aft} = 0$ or 1 to the contribution of each volatile species to the total pebble accretion rate $\Dot{M}_{\rm peb}$. A factor of 0 means that the volatile is not accreted. In this case, 
$\Dot{M}_{\rm peb}$ is lowered by the contribution of that volatile before it is used to determine the growth of the planet during the current time step. Then, we also determine the location of the silicate evaporation front $R_{\rm sil}$ using the temperature and pressure profiles of the envelope. In the next time step, we calculate the total accretion luminosity $L_{\rm tot}$ using $R_{\rm sil}$ and $\Dot{M}_{\rm peb}$, taking into account the factors of the contributions of the corresponding volatiles from the previous time step. If $R_{\rm sil} > r_{\rm core}$ or $\max (T_{\rm env}) > T_{{\rm evap,}Y_{\rm vol}} > T_{\rm rec}$ for any volatile species $Y_{\rm vol}$, latent heat is considered as well. With the total accretion luminosity, the structure equations of the envelope can be integrated. We replace the accretion luminosity calculated by the \texttt{chemcomp} code with $L_{\rm tot}$ when computing the thermal torque while utilizing the envelope module.

Due to the step function-like nature of $T_{\rm rec}$ and the latent heat terms in the expression of $L_{\rm tot}$, the envelope is susceptible to oscillations. In order to limit the magnitude of the oscillations, we damp the rates of change of the latent heat and the factors that govern the accretion of the volatiles. This is done by using both their values calculated \textit{after the integration} of the envelope in the previous time step ($f_{\rm acc,aft}$ and $L_{\rm latent,aft}$) and their initial values used \textit{for the integration} during the previous time step ($f_{\rm acc,int}$ and $L_{\rm latent,int}$). The effective values utilized in the current time step ($f_{\rm acc,cur}$ and $L_{\rm latent,cur}$) are calculated as follows:
\begin{gather}
    L_{\rm latent,cur} = \frac{ L_{\rm latent,aft} + L_{\rm latent,int}}{2}  \\
    f_{\rm acc,cur} = f_{\rm acc,int} + \frac{ f_{\rm acc,aft} - f_{\rm acc,int} }{10}.
    \label{eq: smoothing volatile accretion}
\end{gather}
The current values become $f_{\rm acc,int}$ and $L_{\rm latent,int}$ in the next time step and $f_{\rm acc,aft}$ and $L_{\rm latent,aft}$ are determined after the integration of the envelope. In this way, the values of the factors that govern the accretion of the volatiles are not limited to 0 and 1, but can take any value between these two numbers.

Despite of these efforts to dampen the oscillations, the recycling temperature $T_{\rm rec}$ can still undergo fluctuations for short periods of time during the simulations. This is expected, considering the 
step function-like
dependence of $T_{\rm rec}$ on the relative entropy. It does not pose a problem however, because $T_{\rm rec}$ is not a physical property and is only defined to enable a simple treatment of the recycling of the volatiles. The physical properties of the envelope, such as the temperature, pressure, and density profiles of the envelope, as well as the volatile accretion rates onto the planet, are not subject to fluctuations. Therefore, the present model is capable of simulating the envelopes of terrestrial planets with reasonable accuracy.

\subsection{Comparison to Johansen et al. 2021} \label{sect: reproduction johansen+2021}

\begin{figure}[]
    \centering
    \resizebox{\hsize}{!}{\includegraphics{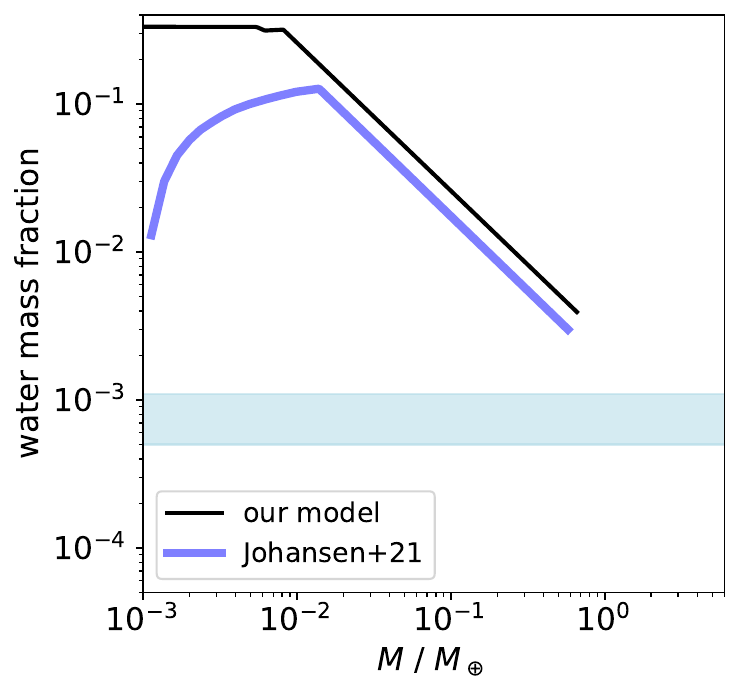}}
    \caption{Planetary water mass fraction of an Earth analog as a function of planetary mass. The light-blue region corresponds to the estimated total water mass fraction of the Earth. The final mass is less than 1 $M_\oplus$ because it is assumed that the proto-Earth subsequently collides with Theia \citep[e.g., ][]{quarles+lissauer2015}. As a result, the proto-Earth gains mass and probably loses some of its water through the impact. The data corresponding to the model of \citet{johansen+2021} were taken from their Fig. 7 for the case that the water ice content of the pebbles in their disk is 35\%. For our model, we used the simple envelope opacity model with $\kappa_{{\rm env},0}$ = 0.1 cm$^2$ g$^{-1}$ and set the pebble surface density to the gas surface density multiplied with $\xi_{\rm flux} = 0.0036$, similar to \citet{johansen+2021}. This corresponds to the {\it "simple opacity"}-models shown in Fig. \ref{fig: johopac}. In addition, we changed the input parameters of the disk and the host star to mimic the conditions in the model of \citet{johansen+2021}. They are listed in Table \ref{tab: johansen comparison}. The evaporation of the pebbles in the disk is neglected.}
    \label{fig: comparison johansen}
\end{figure}

\begin{table}[h]
    \caption{Simulation parameters used for the comparison to Johansen et al. 2021.}
    \centering
    \begin{tabular}{c|c|c}
    \hline\hline
    Parameter & Value & Meaning  \\
    \hline
    \multicolumn{3}{c}{Planet} \\
    \hline
    $a_{p,0}$ & 1.6 AU & initial position \\
    $t_0$ & 0.1 Myr & starting time \\
    $\rho_{\rm core}$ & 5.5 g\ cm$^{-3}$ & density of the planetary core \\
    \hline
    \multicolumn{3}{c}{Grid} \\
    \hline
    $r_{\rm in}$ & 0.1 AU & inner edge \\
    $r_{\rm out}$ & 1000 AU & outer edge \\
    $N_{\rm Grid}$ & 500 & Number of grid-cells \\
    \hline
    \multicolumn{3}{c}{Disk} \\
    \hline
    $\alpha$ & $1\times10^{-4}$ & viscous turbulence parameter \\
    $\alpha_z$ & $1 \times 10^{-4}$ & vertical mixing parameter \\
    $M_*$ & $1\ M_\odot$ & host star mass \\
    $L_*$ & $1\ L_\odot$ & host star luminosity \\
    \text{[Fe/H]} & 0.0 & host star metallicity \\
    $M_0$ & $0.042\ M_\odot$ & initial disk mass \\
    $R_0$ & 150 AU & initial disk radius \\
    $t_{\rm evap}$ & 3 Myr & disk lifetime \\
    $\epsilon_0$ & 2\% & initial solid to gas ratio \\
    $u_{\rm frag}$ & 1 m s$^{-1}$ & solid fragmentation velocity \\
    \hline
    \multicolumn{3}{c}{Integration} \\
    \hline
    $\Delta t$ & 10 yr & integration time step \\
    \hline
    \end{tabular}
    \label{tab: johansen comparison}
\end{table}

In the simulations of \citet{johansen+2021}, their Earth analog accreted water-rich pebbles until it reached a mass of ${\sim}\ 0.02\ M_\oplus$ in the case where the water mass fraction of the disk was set to 35\%. Afterwards, water was recycled back into the disk, causing water ice accretion to cease. The final planetary water mass fraction was approximately 0.003.

In their protoplanetary disk model, \citet{johansen+2021} expressed the inward pebble flux as a constant mass fraction ($\xi_{\rm flux} = 0.0036$) of the gas flux from the outer disk. This resulted in a different evolution of the solid surface density compared to the disk model utilized by the \texttt{chemcomp} code, 
where we use the model of \citet{birnstiel+2012} which means that the pebble growth and drift is an equilibrium between coagulation, fragmentation, and drift.
Since the pebbles had relatively small Stokes numbers in the disk model of \citet{johansen+2021}, the pebble drift speed was low enough that $\xi_{\rm flux}$ approximately represented the ratio of the pebble surface density relative to the surface density of the gas.

The final water content of the planet is also influenced by external parameters, such as the thermodynamic conditions of the disk at the location of the planet, the chemical composition of the disk, and the incoming flux of pebbles. %(as it will be shown in the following sections). 
We can chose these parameters in a way that our model produces similar results as the model of \citet{johansen+2021} when
adopting the assumption that the inward pebble flux is a constant mass fraction ($\xi_{\rm flux} = 0.0036$) of the gas flux.

We show a comparison of our model and the model used by \citet{johansen+2021} in Fig. \ref{fig: comparison johansen}.
Contrary to the model of \citet{johansen+2021}, in which the planetary water mass fraction increases while accreting water-rich pebbles, the chemical composition of the planetary embryo in our model is similar to the composition of the solids in its vicinity. 
Thus, the water mass fraction of our planet does not change significantly during the accretion of water-rich pebbles. This leads to a small enhancement of the final planetary water content compared to the model of \citet{johansen+2021}.
On the whole, our model can reproduce the results of \citet{johansen+2021} with reasonable accuracy, even though the underlying models of the protoplanetary disk are different\footnote{For example, \citet{johansen+2021} consider the accretion of planetesimals in their model, which are not present in our model.}.

\section{Self-consistent envelope opacity model} 

\subsection{Dust and pebbles in the planetary envelope} \label{sect: self-consistent opacity}

Following \citet{brouwers+2021}, the pebble size $R_{\rm peb}$ is given by
\begin{gather}
    R_{\rm peb} = \min (  R_{\rm coal}, R_{\rm vlim} ),
\end{gather}
where $R_{\rm coal}$ is the pebble size regulated by the collision and the sedimentation timescale, and $R_{\rm vlim}$ is the pebble size governed by fragmentation and/or erosion. The dust size is $R_{\rm dust} = 1\ \mu$m and remains constant. Dust is produced in pebble-pebble collisions and in high-velocity pebble-dust encounters (i.e., erosion), while it is lost by sticking encounters with slower pebbles (so-called dust sweep-up). This results in a local steady state between dust production via fragmentation/erosion and dust loss via sweep-up.

Both the dust and the pebbles feel the particle size dependent drag force induced by the gas background in the envelope, effectively slowing them down. \citet{brouwers+2021} adopted a simple, continuous two-regime approximation to express the drag force. The transition from the Epstein regime (i.e., free molecular flow) to the Stokes regime (i.e., continuum flow) occurs when the Knudsen number of the particles becomes 4/9. Thus, particles of the size $R_{\rm s}$ are within the Epstein flow regime if
\begin{gather}
    \frac{l_{\rm mfp}}{R_{\rm s}} < \frac{4}{9},
    \label{eq: drag regime envelope}
\end{gather}
where $l_{\rm mfp}$ is the mean free path of the molecules and $R_{\rm s}$ can be either the size of the pebbles or the size of the dust grains.
If Eq. \ref{eq: drag regime envelope} is not satisfied, the particles are within the Stokes regime.

In the envelope, the dust and the pebbles sink at a speed close to their terminal velocities $v_{\rm fall}$. The terminal velocity of a particle can be expressed as \citep{weidenschilling1977}
\begin{gather}
    v_{\rm fall} = \frac{g R_{\rm s} \rho_{\rm env,\bullet}}{\rho_{\rm env} v_{\rm th}} \max \left( \frac{4 R_{\rm s}}{9 l_{\rm mfp}}, 1 \right),
\end{gather}
where $g = G M / r_{\rm env}^2$ is the gravitational acceleration, $\rho_{\rm env,\bullet} = 3.2$ g cm$^{-3}$ is the density of a solid particle, and the mean thermal velocity of the gas $v_{\rm th}$ is given by
\begin{gather}
    v_{\rm th} = \sqrt{ \frac{8 k_B T_{\rm env}}{\pi \mu m_{\rm p}} }.
\end{gather}
The velocity of the downward flow of the gas in the envelope $v_{\rm gas}$ can be approximated from the mass conservation of the gas accretion rate onto the planet $\Dot{M}_{\rm gas}$ as
\begin{gather}
    v_{\rm gas} = \frac{\Dot{M}_{\rm gas}}{4 \pi r_{\rm env}^2 \rho_{\rm env}}.
\end{gather}
Due to their small size, the sedimentation velocity of the dust grains is dominated by the contribution of the gas flow \citep{mordasini2014}. In the accretion model used by the \texttt{chemcomp} code, the planet does not accrete gas before it has reached the pebble isolation mass, causing the velocity of the gas flow $v_{\rm gas}$ to vanish during pebble accretion in our model.

By assuming that the total velocities of the pebbles are dominated by the terminal component, the maximum size to which the pebbles can grow as they sediment can be expressed as \citep{brouwers+2021}
\begin{gather}
    R_{\rm coal} = \left\{\begin{array}{ll} \left( \frac{3 x_{\rm R} H_{\rm env} \Dot{M}_{\rm peb} v_{\rm th} \rho_{\rm env}}{16 \pi G M \rho_{\rm env,\bullet}^2} \right) ^{1/2}, & \text{Epstein}\\
    \frac{3}{4} \left( \frac{x_{\rm R} H_{\rm env} \Dot{M}_{\rm peb} v_{\rm th} l_{\rm mfp} \rho_{\rm env}}{\pi G M \rho_{\rm env,\bullet}^2} \right)^{1/3}, & \text{Stokes} \end{array}\right. ,
    \label{eq: maximum size pebbles can grow to during their sedimentation}
\end{gather}
where $H_{\rm env} = k_B T_{\rm env} / (\mu g)$ is the envelope scale height and $x_{\rm R} = v_{\rm col}/v_{\rm fall} = 1/3$ is the ratio of the pebble collision velocity $v_{\rm col}$ and the terminal velocity. The maximum pebble size $R_{\rm coal}$ linearly depends on the pebble accretion rate onto the planet $\Dot{M}_{\rm peb}$, while it remains constant across changes in the planetary mass \citep{brouwers+2021}.

At higher velocities, the pebbles can fragment upon collision or experience significant amounts of erosion. Both processes can be combined to obtain a limit for the terminal velocity of the pebbles:
\begin{gather}
    v_{\rm lim} = \min \left( v_{\rm erosion}, \frac{v_{\rm frag}}{x_{\rm R}} \right).
\end{gather}
The erosion velocity $v_{\rm erosion}$ is given by \citep{schräpler+2018}
\begin{gather}
    v_{\rm erosion} = 2.4 \text{m s$^{-1}$} \left( \frac{R_{\rm dust}}{1\ \mu\text{m}} \right) ^{1/1.62},
\end{gather}
which means that $v_{\rm erosion} = 2.4\ \text{m s$^{-1}$}$ in the model of \citet{brouwers+2021}. In addition, \citet{brouwers+2021} set the fragmentation threshold in the envelope to $v_{\rm frag} = 0.8$ m s$^{-1}$, causing the limiting velocity to be $v_{\rm lim} = 2.4$ m s$^{-1}$ throughout the entire envelope. The velocity-limited pebble size can then be expressed as \citep{brouwers+2021}
\begin{gather}
    R_{\rm vlim} = \left\{\begin{array}{ll} \frac{\rho_{\rm env} v_{\rm th} v_{\rm lim}}{g \rho_{\rm env,\bullet}}, & \text{Epstein}\\
    \frac{3}{2} \left( \frac{\rho_{\rm env} v_{\rm th} l_{\rm mfp} v_{\rm lim}}{g \rho_{\rm env,\bullet}} \right)^{1/2}, & \text{Stokes} \end{array}\right. .
    \label{eq: velocity-limited pebble size}
\end{gather}
Contrary to the pebble size regulated by collision and sedimentation $R_{\rm coal}$, the velocity-limited pebble size $R_{\rm vlim}$ scales positively with the planetary mass. Typically, the pebble sizes in the envelopes of lower-mass planets are velocity-limited, whereas the pebbles sizes in the envelopes of more massive planets are only limited by their rate of growth.

The amount of dust harbored by the envelope is governed by the fractional dust production efficiency $F$, which can take values between 0 (no dust-production, i.e., no contribution of the dust to the total envelope opacity) and 1 (erosion-limited regime, i.e., all growth beyond $R_{\rm vlim}$ is transferred into dust). Physically, $F$ is a proxy for the number of pebble-pebble collisions needed to completely grind down a pebble \citep{brouwers+2021} and connects the dust density in the envelope $\rho_{\rm env,dust}$ to the pebble density $\rho_{\rm env,peb}$ as
\begin{gather}
    \frac{\rho_{\rm env,dust}}{\rho_{\rm env,peb}} = F x_{\rm R}.
\end{gather}
\citet{brouwers+2021} used a default value of $F = 0.1$ which causes the contribution of the dust to the total opacity to be comparable to the contribution of the pebbles. 

\subsection{Contributions to the total opacity} \label{sect: opacity contribution}

In general, the Rosseland mean opacity of a population of solids $\kappa_{\rm s}$ in the planetary envelope can be expressed as \citep{movshovitz+podolak2008}
\begin{gather}
    \kappa_{\rm s} = \frac{3 Q_{\rm eff} \rho_{\rm env,s}}{4 \rho_{\rm env,\bullet} R_{\rm s} \rho_{\rm env}},
    \label{eq: rosseland mean opacity}
\end{gather}
where $\rho_{\rm env,s}$ is either the dust or the pebble volume density in the envelope and the extinction coefficient $Q_{\rm eff}$ can be approximated as
\begin{gather}
    Q_{\rm eff} \simeq \min \left( \frac{0.6 \pi R_{\rm s}}{\lambda_{\rm peak}}, 2 \right).
\end{gather}
The peak wavelength of the photons $\lambda_{\rm peak}$ emitted by the local gas can be found through Wien's Law:
\begin{gather}
    \lambda_{\rm peak} = \frac{0.29 \text{ cm K}}{T_{\rm env}}.
\end{gather}
\citet{brouwers+2021} neglected the species-dependency of the solid opacity \citep[e.g., ][]{bitsch+savvidou2021} in their formulation. However, they did account for the sublimation of silicates at high temperatures by omitting the opacities of the dust and the pebbles in the layers of the envelope where the local temperature is higher than the silicate sublimation temperature (which was set to 2500 K in their work).

\paragraph{Pebble opacity}

If the pebbles in the envelope tend to have relatively low velocities, they grow efficiently through sticking collisions and their typical size $R_{\rm coal}$ can be found through Eq. \ref{eq: maximum size pebbles can grow to during their sedimentation}. By plugging $R_{\rm coal}$ in the expression of the Rosseland mean opacity (Eq. \ref{eq: rosseland mean opacity}), the opacity of the pebbles in the growth limited case is given by the following equation in both the Epstein and the Stokes drag regimes \citep{mordasini2014, brouwers+2021}:
\begin{gather}
    \kappa_{\rm coal} = \frac{Q_{\rm eff}}{x_{\rm R} H_{\rm env} \rho_{\rm env}}.
\end{gather}

If the terminal velocity of the pebbles is higher than the limiting velocity $v_{\rm lim}$, their typical size is not greater than $R_{\rm vlim}$ (Eq. \ref{eq: velocity-limited pebble size}). Plugging $R_{\rm vlim}$ in the expression of the Rosseland mean opacity (Eq. \ref{eq: rosseland mean opacity}) yields the pebble opacity in the velocity limited case \citep{brouwers+2021}:
\begin{gather}
    \kappa_{\rm vlim} = \left\{\begin{array}{ll} \frac{3 Q_{\rm eff} \Dot{M}_{\rm peb} g}{16 \pi r_{\rm env}^2 \rho_{\rm env}^2 v_{\rm th} v_{\rm lim} (v_{\rm lim} + v_{\rm gas})}, & \text{Epstein}\\
   \frac{Q_{\rm eff} \Dot{M}_{\rm peb}}{8 \pi r_{\rm env}^2 (v_{\rm lim} + v_{\rm gas})} \left( \frac{g}{\rho_{\rm env,\bullet} \rho_{\rm env}^3 v_{\rm th} v_{\rm lim} l_{\rm mfp}} \right) ^{1/2}, & \text{Stokes} \end{array}\right. .
\end{gather}

With $\kappa_{\rm coal}$ and $\kappa_{\rm vlim}$, the effective pebble opacity $\kappa_{\rm peb}$ can be found through
\begin{gather}
    \kappa_{\rm peb} = \left\{\begin{array}{ll} \kappa_{\rm vlim}, & R_{\rm vlim} < R_{\rm coal}\\
   \kappa_{\rm coal}, & R_{\rm vlim} > R_{\rm coal} \end{array}\right. .
\end{gather}
This equation has to be evaluated locally at every depth in the envelope.

\paragraph{Dust opacity}

In the steady state between dust production via fragmentation/erosion and dust loss via sweep-up through pebbles, the dust opacity is given by \citep{brouwers+2021}
\begin{gather}
    \kappa_{\rm dust} = \kappa_{\rm peb} \frac{Q_{\rm eff,dust}}{Q_{\rm eff,peb}} \frac{R_{\rm peb}}{R_{\rm dust}} F x_{\rm R}.
    \label{eq: dust opacity}
\end{gather}
The dust opacity linearly depends on both the pebble opacity $\kappa_{\rm peb}$ and the dust production efficiency parameter $F$. When using the default value of \citet{brouwers+2021}, i.e., $F = 0.1$, the dust opacity is typically comparable to the pebble opacity.

\paragraph{Gas opacity}

\citet{brouwers+2021} did not model the gas opacity in detail, but adopted the simple analytic scaling of the molecular opacity by \citet{bell+lin1994}:
\begin{gather}
    \kappa_{\rm gas} = 10^{-8} \rho_{\rm env}^{2/3} T_{\rm env}^3 \text{ cm$^2$ g$^{-1}$}.
\end{gather}
At depths in the envelope where the silicate particles sublimate, the gas opacity is the only contribution to the total opacity.
The total envelope opacity can then be found using Eq. \ref{eq: envelope opacity brouwers}.

\subsection{Implementation} \label{sect: implementation opacity}

When we use the Brouwers model in our simulations, it is always used in tandem with the envelope model discussed before. In this case, the simple envelope opacity law (Eq. \ref{eq: envelope opacity johansen}) is replaced by the expression of the Brouwers opacity (Eq. \ref{eq: envelope opacity brouwers}). The necessary properties, such as the typical pebble sizes in the different growth regimes ($R_{\rm coal}$, $R_{\rm vlim}$, as well as $R_{\rm peb}$) and the three contributions to the total opacity ($\kappa_{\rm peb}$, $\kappa_{\rm dust}$, and $\kappa_{\rm gas}$), are then included into the integration routine that solves the structure equations of the envelope. Therefore, these properties are evaluated at each time step over the whole radial range of the envelope using the local thermal conditions. This significantly increases the computation time.

In order for the envelope model and the envelope opacity model to be consistent with each other, we do not employ a constant silicate sublimation front at $T_{\rm env} = 2500$ K, as it was done by \citet{brouwers+2021}. Instead, the silicate sublimation front is located at the depth where the saturated vapor pressure of forsterite (Eq. \ref{eq: silicate sublimation front envelope}) becomes greater than the local pressure of the envelope (i.e., $R_{\rm sil}$). Thus, the total envelope opacity $\kappa_{\rm env}$ does only consist of the gas opacity $\kappa_{\rm gas}$ in regions where $r_{\rm env} < R_{\rm sil}$.

\subsection{Default parameters vs. lower limit} \label{sect: default parameters vs. lower limit}

\begin{figure}[]
    \centering
    \resizebox{\hsize}{!}{\includegraphics{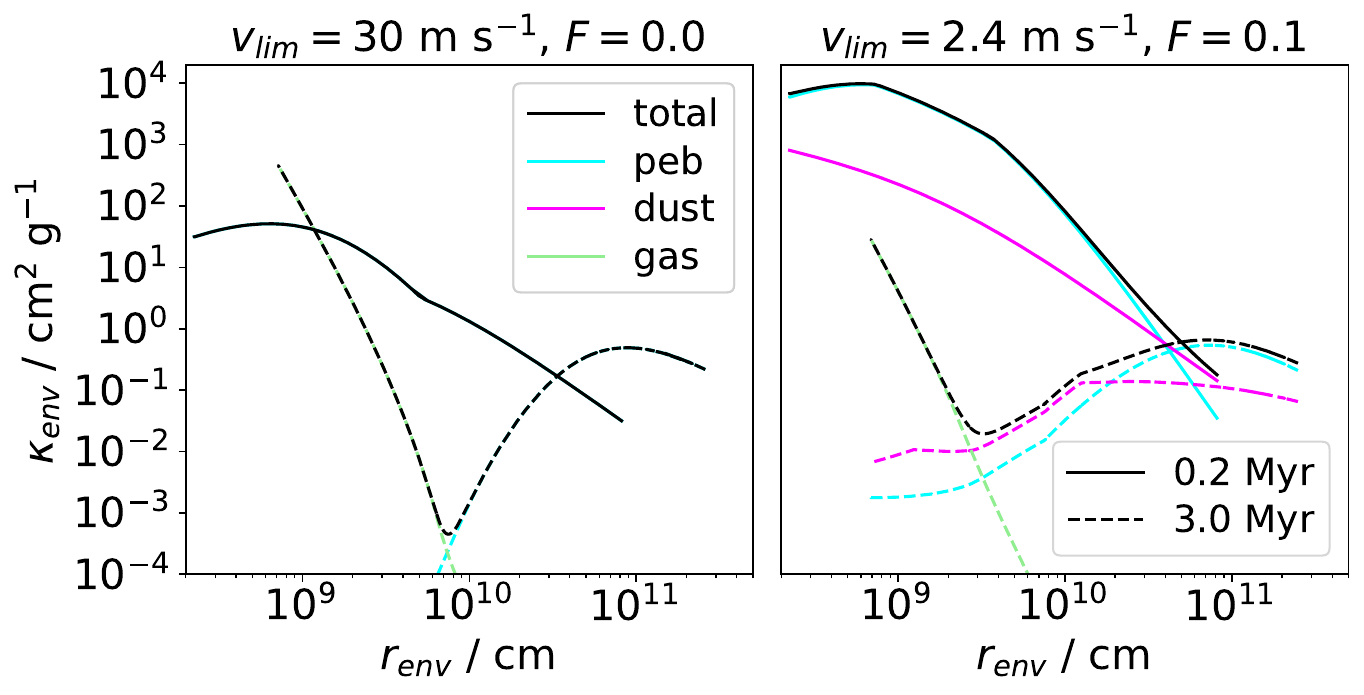}}
    \caption{Brouwers opacity as a function of the distance from the center of the planetary core at an early and a late stage in the growth of the planet. \textit{Left panel:} Estimate of the lower limit of the envelope opacity ($v_{\rm lim} = 30.0$ m s$^{-1}$, $F = 0.0$). \textit{Right panel:} Envelope opacity that occurs when the default values of \citet{brouwers+2021} are used ($v_{\rm lim} = 2.4$ m s$^{-1}$, $F = 0.1$). The contributions of the gas, the dust, and the pebbles to the total opacities are featured as well. The migration of the planet is suppressed, the growth of the planet initiates at $t_0 = 0.1$ Myr, and the disk lifetime is 3 Myr. The incoming pebble flux is fixed to $0.6\ M_\oplus$ Myr$^{-1}$, but can still be lowered by volatile recycling in the planetary envelope. 
    Apart from that, we use our standard parameters (see Table \ref{tab: simulation parameters}). After 0.2 Myr (3.0 Myr), the planets have a mass of about 0.05 $M_\oplus$ (1.5 $M_\oplus$).}
    \label{fig: brouwers opacity contributions}
\end{figure}

In Fig. \ref{fig: brouwers opacity contributions}, we show the Brouwers opacity calculated using the model by \citet{brouwers+2021}, including the individual contributions of the gas, the dust, and the pebbles. We consider both the default values of \citet{brouwers+2021} (i.e., $v_{\rm lim} = 2.4$ m s$^{-1}$, $F = 0.1$), as well as an estimate of the lower limit of the envelope opacity assuming a limiting velocity corresponding to a solid fragmentation velocity of 10 m s$^{-1}$ and no dust production in the envelope (i.e., $v_{\rm lim} = 30$ m s$^{-1}$, $F = 0.0$). In order to exclude external influences, we suppress the migration of the planet and adopt a constant incoming pebble flux of $\Dot{M}_{\rm peb} = 0.6\ M_\oplus$ Myr$^{-1}$ onto the planet. The planet resides at a disk temperature of 100 K.

When using the default values, the envelope opacity is initially dominated by the contribution of the pebbles. Only near the outer edge of the envelope, the dust opacity becomes larger than the pebble opacity. Since the envelope is cold at the beginning of the planet's growth, the gas opacity does not play a role at that time. At the end of the planet's growth, the envelope opacity has become significantly smaller. This is mainly caused by a strong decrease in the pebble opacity throughout the entire envelope, but the dust opacity has decreased as well. The inner, hot parts of the envelope are now dominated by the gas opacity, which takes over at high temperatures and increases strongly with increasing depth. Nevertheless, the gas opacity is still not as high as the initial pebble opacity, even in the inner envelope.

When we consider the lower limit of the envelope opacity by increasing the limiting velocity and suppressing the dust production in the envelope, the envelope opacity is lower than when using the default parameters. Firstly, this is due to the vanishing dust opacity caused by the absence of dust in the envelope. Secondly, the higher limiting velocity allows the pebbles to grow to larger sizes in the velocity-limited regime, thus reducing the pebble opacity. In the early stages of the planet's growth, the latter is the dominating effect.

At larger depths, the pebble opacity becomes significantly lower in the case of the lower limit than when using the default parameters. This is an indirect consequence of the envelope opacity which is lower overall when considering the lower limit. A lower envelope opacity leads to a colder envelope, which enables the planet to accrete more icy volatiles. Since we only fix the incoming pebble flux onto the planet, the effective pebble accretion rate can still be lowered by the recycling of volatiles in the envelope. Therefore, the planet is more massive when considering the lower limit of the envelope opacity than the planet using the default parameters. The higher mass of the planet when considering the lower limit is also the reason why its corresponding gas opacity is higher at the same distances from the center of the planet than when using the default parameters.

In both cases, the envelope opacity generally decreases with increasing planetary mass. For details, see \citet{brouwers+2021}.

\section{Modeling the TRAPPIST-1 planets neglecting volatile recycling}  \label{sect: no volatile recycling}

\begin{figure*}[]
    \centering
    \resizebox{\hsize}{!}{\includegraphics{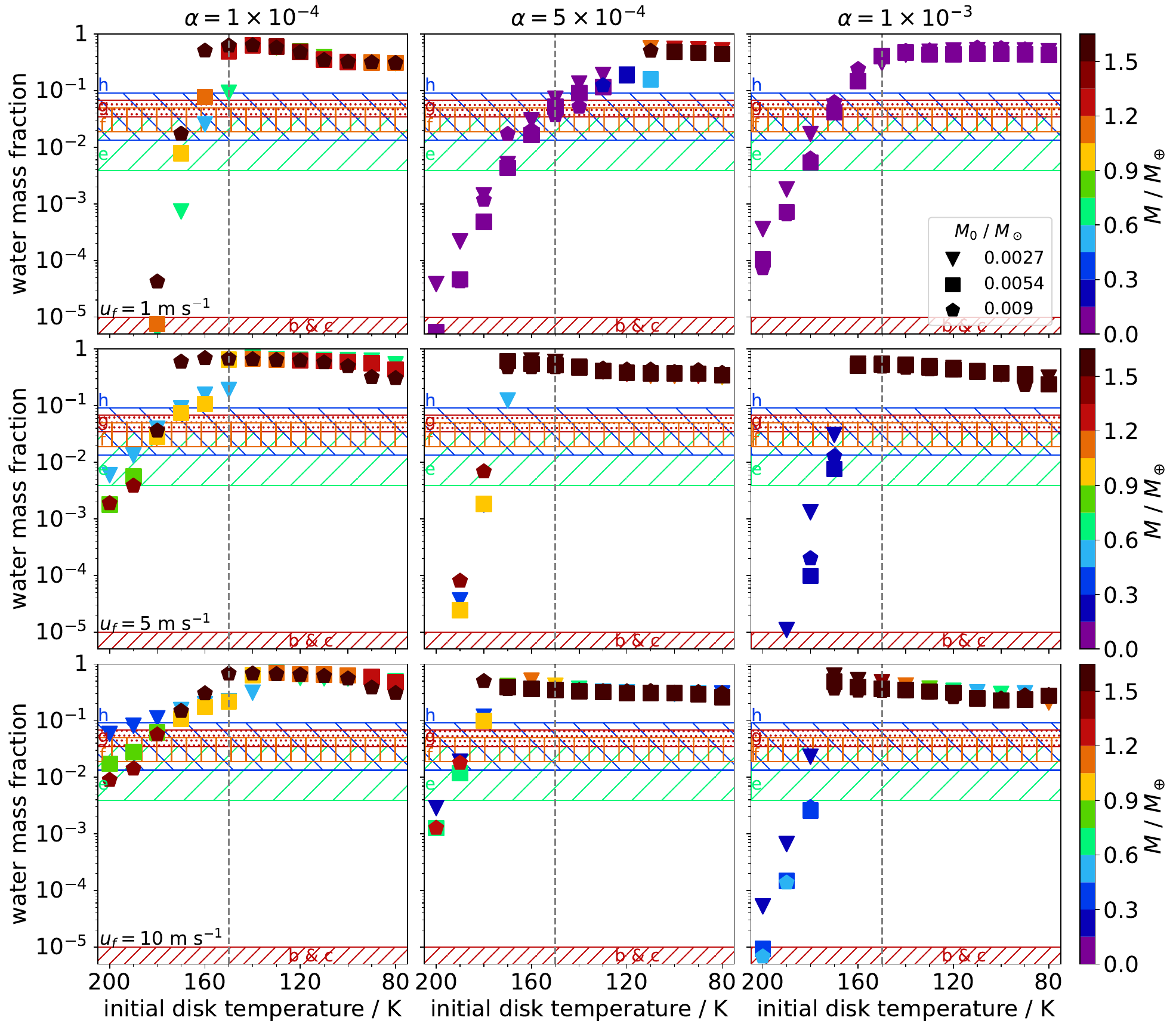}}
    \caption{Same as Fig. \ref{fig: ENVTOT evap}, but neglecting the recycling of volatiles in the planetary envelope.
    The water mass fractions of the planets can exceed the initial maximum water content of the solids in our disk model (i.e., ${\sim}$35\%) if they accrete significant amounts of their final mass near the water-evaporation line. At that location, the water content of the solids is much higher due to local recondensation of water vapor \citep[e.g., ][]{drazkowska+alibert2017, ros+2019}, see also \citet{schneider+bitsch2021a}. This is similar to the effect for iron grains in the inner disk, which can explain the formation of super-Mercuries \citep{mah+bitsch2023}.}
    \label{fig: NOENV evap}
\end{figure*}

Here, we adopt the classical model of planetary pebble accretion, where the influence of the planetary envelope is neglected (i.e., atmospheric sublimation and recycling of pebbles are neglected). In the case of a nonmigrating planet, this would mean that the planet's water content is given by the abundance of water in the solids surrounding the planet in the disk. However, the planets are free to migrate in our simulations. Since the water mass fraction of the solids in the disk either vanishes interior to the water-evaporation line or lies between ${\sim}$20\% and ${\sim}$30\% exterior to the water-evaporation line, the planet must cross the water-evaporation line during its growth to be able to realize the moderate water mass fractions estimated by e.g., , \citet{raymond+2022} for the outer planets of the TRAPPIST-1 system \citep[e.g., ][]{bitsch+2019b, schoonenberg+2019}.
The resulting planetary water mass fractions are shown in Fig. \ref{fig: NOENV evap}.

The pebble-accretion model which neglects the recycling of volatiles seems to have difficulties in explaining the water contents and masses of the outer TRAPPIST-1 planets. Although pebble evaporation in the disk in combination with the efficient heating-torque flattens the transition form water-rich planets to water-poor planets, increasing the probability of a planet having an intermediate water abundance (as it is inferred for the outer TRAPPIST-1 planets), the planets that have moderate water mass fractions tend to be too low in mass in most of our disks.
Planets matching both the water mass fraction and the mass of one of the heavier of the outer TRAPPIST-1 planets, that is, of the planets f and g, are rare and constrained to very specific initial positions in massive disks with a low $\alpha$-viscosity.
We therefore conclude that it is rather unlikely that the pebble-accretion model, which does not account for the recycling of volatiles, can explain the estimated water contents and masses of {\it all} the outer TRAPPIST-1 planets in a consistent manner.

\end{appendix}
\end{document}